\documentclass[preprint,tightenlines,aps,prd,showpacs,superscriptaddress,
nofootinbib,preprintnumbers, 11pt]{revtex4-1}

\usepackage[utf8]{inputenc}
\usepackage[english]{babel}
\usepackage{graphicx}
\usepackage{epstopdf}

\usepackage{amsmath}
\usepackage{braket}
\usepackage{bbm}

\usepackage{xcolor}
\definecolor{blue}{rgb}{0.25, 0.41, 0.88}
\definecolor{red}{rgb}{0.86, 0.08, 0.24}
\definecolor{green}{rgb}{0.2, 0.60, 0.2}
\definecolor{magenta}{rgb}{0.7, 0, 0.5}
\definecolor{cyan}{rgb}{0.5, 0.8, 0.8}
\definecolor{yellow}{HTML}{CCCC00}
\definecolor{violet}{HTML}{7F00FF}
\definecolor{orange}{HTML}{FF8000}
\definecolor{gray}{HTML}{888888}

\usepackage{hyperref}
\begin{document}

\title{Electric structure of shallow \textit{D}-wave states in Halo 
EFT}

\author{J. Braun}
\email{braun@theorie.ikp.physik.tu-darmstadt.de}
\affiliation{Institut für Kernphysik, Technische Universität Darmstadt,
64289 Darmstadt, Germany}

\author{W. Elkamhawy}
\email{elkamhawy@theorie.ikp.physik.tu-darmstadt.de}
\affiliation{Institut für Kernphysik, Technische Universität Darmstadt,
64289 Darmstadt, Germany}

\author{R. Roth}
\email{robert.roth@physik.tu-darmstadt.de}
\affiliation{Institut für Kernphysik, Technische Universität Darmstadt,
64289 Darmstadt, Germany}

\author{H.-W. Hammer}
\email{Hans-Werner.Hammer@physik.tu-darmstadt.de}
\affiliation{Institut für Kernphysik, Technische Universität Darmstadt,
64289 Darmstadt, Germany}
\affiliation{ExtreMe Matter Institute EMMI, GSI Helmholtzzentrum
für Schwerionenforschung GmbH, 64291 Darmstadt, Germany}

\date{\today}

\begin{abstract}
We compute the electric form factors of one-neutron halo nuclei with shallow 
$D$-wave states up to next-to-leading order and 
the E2 transition from the $S$-wave to the $D$-wave state up to leading order 
in Halo Effective Field Theory (Halo EFT). The relevant degrees of freedom 
are the core and the halo neutron. The EFT expansion is carried out in powers 
of $R_{core}/R_{halo}$, where $R_{core}$ and $R_{halo}$ denote the length scales
of the core and the halo, respectively.
We propose a power counting scenario for weakly-bound 
states in one-neutron Halo EFT and discuss its implications for higher partial 
waves in terms of universality. The scenario is applied to the 
$\frac{5}{2}^+$ first excited state and the $\frac{1}{2}^+$ ground state of 
$^{15}$C. We obtain several universal correlations between electric observables 
and use data for the E2 transition $\frac{5}{2}^+\to 
\frac{1}{2}^+$ together with \textit{ab initio} results from the No-Core Shell 
Model to predict the quadrupole moment.
\end{abstract}

\maketitle

\section{Introduction}
The quantitative description of halo nuclei in Halo Effective Field
Theory (Halo EFT) provides insights into their universal properties.
Halo nuclei consist of a tightly bound 
core nucleus surrounded by one or more weakly bound
nucleons~\cite{jensen2004structure,Riisager:2012it}.
This separation of scales can be captured in terms of the core
length scale, $R_{core}$, and the halo scale, $R_{halo}$, with
$R_{halo} \gg R_{core}$.
Halo EFT exploits this separation of scales to describe halo
nuclei~\cite{bertulani2002effective,bedaque2003narrow}. In this 
approach, the relevant degrees of freedom are the core and the halo nucleons. 
Halo EFT is complementary to \textit{ab initio} methods that have 
difficulties describing weakly-bound states and provides
a useful tool to identify 
universal correlations between observables. For recent reviews of
Halo EFT see Refs.~\cite{ando2016hypernuclei,rupak2016radiative,Hammer:2017tjm}.

The Halo EFT formalism has been successfully used to study 
various reactions and properties of halo-like systems. Some early
examples in the 
strong sector include the $n\alpha$ resonance in
$^5$He~\cite{bertulani2002effective,bedaque2003narrow}
the $\alpha\alpha$ resonance in ${}^8$Be~\cite{higa2008alphaalpha}
and universal properties, matter form 
factors and radii of two-neutron halo
nuclei with predominantly $S$-wave~\cite{canham2008universal,canham2010range}
and $P$-wave interactions \cite{Rotureau:2012yu,Ji:2014wta}.
Due to the importance of higher partial waves in
halo nuclei, different 
power counting schemes are conceivable that have a varying number of
fine tuned
parameters~\cite{bertulani2002effective,bedaque2003narrow}. From
naturalness assumptions, one expects a lower number of fine tunings
to be more likely to occur in nature. However, the level of fine 
tuning depends strongly on the details of the considered system and has to be 
verified and adjusted to data.

In Halo EFT, electromagnetic interactions can be straightforwardly included via 
minimal substitution in the Lagrangian, and relevant electromagnetic currents 
can be added. Some applications to one-neutron halos,
which we consider here,
are the calculation of electric 
properties of $^{11}$Be~\cite{hammer2011electric}, $^{15}$C 
\cite{fernando2015electromagnetic},
radiative neutron capture on 
${}^7$Li~\cite{rupak2011model,zhang2014combiningLi}
and $^{14}$C~\cite{rupak2012radiative},
the ground state structure of 
${}^{19}$C~\cite{acharya201319},
and the electromagnetic properties of $^{17}$C~\cite{Braun:2018vez}. 
The parameters needed as input in Halo EFT can be either taken from experiment 
or from \textit{ab initio} 
calculations~\cite{Hagen:2013xga,zhang2014combiningBe,zhang2014combiningLi},
which shows the 
versatility and complementary character of Halo EFT.

Electric properties provide a unique window on the structure and dynamics of 
one-neutron halo nuclei. 
In this work, we consider $^{15}$C as an example
and follow the approach presented in 
Ref.~\cite{hammer2011electric}, where electric properties of ${}^{11}$Be are 
calculated using Halo EFT. $^{15}$C also has two bound states. The 
$\frac{1}{2}^+$ ground state of ${}^{15}$C is predominantly an $S$-wave bound 
state, and the $\frac{5}{2}^+$ first excited state predominantly a $D$-wave 
bound state.
Therefore, we focus on the extension to partial waves beyond the $P$-wave, 
in general, and especially on the extension to $D$-wave states.  We 
include the strong $D$-wave interaction by introducing a new dimer field and 
compute the E2 transition strength and electric form factors.
In the context of the strong
$d+t \leftrightarrow n+\alpha$ reaction, $D$-wave states were 
also investigated in Ref.~\cite{brown2014field}. We use a similar approach 
for dressing the $D$-wave propagator, but a different regularization scheme. 
This entails a different power counting scheme as will be discussed 
in more detail in the next section.

The paper is organized as follows: After writing down the non-relativistic 
Lagrangian for the $S$- and $D$-wave case in Section 
\ref{sec:Halo_EFT_Formalism}, we dress the $S$- and $D$-wave propagators. As 
regularization scheme, a momentum cutoff is employed to identify all 
divergences. For practical calculations, the power divergence 
subtraction scheme~\cite{kaplan1998new, kaplan1998two} is applied
for convenience. Based on our analysis of the divergence structure,
we propose a power counting scenario and discuss its 
implications for higher partial wave bound states in terms of universality.
In Ref.~\cite{Braun:2018vez}, the same power counting as in this paper is applied in order
to describe shallow $D$-wave bound states in $^{17}$C.
After the inclusion of electric interactions in our theory, the B(E2) 
transition strength between the $S$- and $D$-wave state as well as electric 
form factors of the $D$-wave state are calculated in Section 
\ref{sec:EM_Observables}.
First, we present general results and correlations for such weakly-bound 
systems and then apply them to the case of $^{15}$C. Eventually, our Halo EFT 
results for $^{15}$C are combined with data for the B(E2) 
transition strength~\cite{ajzenberg1991energy} and \textit{ab initio} results
from the Importance-Truncated No-Core Shell Model (IT-NCSM)~\cite{roth2009importance}. In this 
way, we are able to predict the quadrupole and hexadecapole moments and radii. 
Our findings are then compared to 
correlations~\cite{calci2016sensitivities} which are motivated by
the rotational model of Bohr and Mottelson~\cite{bohr1975nuclear}. In Section 
\ref{sec:conclusion}, we present our conclusions.

\section{Halo EFT formalism}
\label{sec:Halo_EFT_Formalism}
We apply the Halo EFT formalism for the electric properties of
$P$-wave systems developed in Ref.~\cite{hammer2011electric} to
shallow $D$-wave systems.
Since we use our theory to describe
${}^{15}$C which has a shallow $S$-wave state ($J^P=\frac{1}{2}^+$) and
a shallow $D$-wave state ($J^P=\frac{5}{2}^+$),
we also include an $S$-wave state in our theory.

\subsection{Lagrangian}

The relevant degrees of freedom are the core, a bosonic field $c$, and the halo 
neutron, a spinor field $n$.
The strong $S$- and $D$-wave interactions are included through auxiliary 
spinor fields $\sigma$ for the $S$-wave
state and $d$ for the $D$-wave states, respectively.
Note, that we include only one $d$ field in the Lagrangian below.
In principle, there are two $d$ fields for the $\frac{5}{2}^+$ and 
$\frac{3}{2}^+$ states, respectively.
Summing over repeated spin indices, the effective Lagrangian can be written as
\begin{align}
\notag
\mathcal{L} = \ & c^\dagger \left(i \partial_t + \frac{\nabla^2}{2M}\right) c
+ n^\dagger \left(i \partial_t + \frac{\nabla^2}{2m_n}\right) n
+ \sigma_s^\dagger \left[\eta_0 \left(i \partial_t + 
\frac{\nabla^2}{2M_{nc}}\right)
+ \Delta_0 \right] \sigma_s \\[10pt] \notag
& + d_{m}^\dagger \left[c_2 \left(i \partial_t + 
\frac{\nabla^2}{2M_{nc}}\right)^2
+ \eta_2 \left(i \partial_t + \frac{\nabla^2}{2M_{nc}}\right)
+ \Delta_2 \right] d_{m}\\[10pt]
\label{eq:lagrangian}
& - g_0 \left[c^\dagger n_s^\dagger \sigma_s + \sigma_s^\dagger n_s c\right] -
g_2^{J} \left[d^\dagger_{m} \left[n \overset{\leftrightarrow}\nabla^2 c\right]_{J,m}
+ \left[c^\dagger \overset{\leftrightarrow}\nabla^2 n^\dagger \right]_{J,m} 
d_{m} \right] +\ldots \ ,
\end{align}
where $3/2 \leq J \leq 5/2$ denotes the total spin of the $D$-wave state,
$m_n$ is the neutron mass, $M$ the core mass and $M_{nc} = m_n + M$
is the total mass of the $nc$ system. The repeated spin 
indices $s$ and $m$ are summed over according to the Einstein convention.
The power counting for this
Lagrangian depends on the underlying scales and will be discussed
below.
The $S$-wave part of Eq.~(\ref{eq:lagrangian})
contains three coupling constants $g_0$, $\Delta_0$ and
$\eta_0$, while only two of them are linearly independent. In principle, we are
free to choose which constant is set to a fixed value. Here, we choose 
$\eta_0 = \pm 1$ to be a sign which will be fixed by the effective range.
(For an alternative choice, see Ref.~\cite{Griesshammer:2004pe}.)
This part is well known and has been discussed extensively in the
literature on Halo EFT~\cite{Hammer:2017tjm,hammer2011electric,fernando2015electromagnetic}. 
To make the paper self-contained,
the key equations for the interacting propagator
of the $S$-wave state are collected in Appendix~\ref{app:Swave}.
In the following, we focus on the properties of the $D$-wave state.
For the $D$-wave, we include four constants in our Lagrangian, namely $c_2$, 
$\eta_2$, $\Delta_2$ and $g_2$.
However, in this case only three of them are linearly
independent. Again, we are free to choose which constant is set to
a fixed value. Here, we choose $\eta_2 = \pm 1$ to be a sign, but other choices
are possible.
The additional 2nd-order kinetic term with constant $c_2$ is needed to
renormalize the interacting $D$-wave propagator which contains up to
quintic divergences.
Since the core and neutron have different masses, it is convenient to define
the interaction terms with Galilei invariant derivatives
\begin{align}
n \overset{\leftrightarrow}\nabla c = n \ \frac{\left(M 
\overset{\leftarrow}\nabla
- m_n \overset{\rightarrow}\nabla \right)}{M_{nc}} \ c \ ,
\end{align}
where the arrows indicate the direction of their action.
We project on the $J=5/2$ or $3/2$ component of the interaction by defining
\begin{align}
\label{eq:d_tensor_L}
\left[n \overset{\leftrightarrow}\nabla^2 c\right]_{J,m} = \sum_{m_s m_l}
\left(\left. \frac{1}{2} m_s \ 2 m_l \right\vert J m \right) \ n_{m_s}
\sum_{\alpha \beta} \left(\left. 1 \alpha \ 1 \beta \right\vert 2 m_l \right) 
\frac{1}{2}
\left( \overset{\leftrightarrow}\nabla_\alpha 
\overset{\leftrightarrow}\nabla_\beta +
\overset{\leftrightarrow}\nabla_\beta \overset{\leftrightarrow}\nabla_\alpha 
\right) \ c \ ,
\end{align}
where $\alpha$ and $\beta$ are spherical indices and 
$\left(\left. j_1m_1 \ j_2 m_2 \right\vert J m \right)$ 
are Clebsch-Gordan coefficients~\cite{Tanabashi:2018oca} 
coupling $j_1$ and $j_2$ with projections $m_1$ and $m_2$, respectively, to 
$J$ with projection $m$. 

In practice, we calculate $D$-wave observables in Cartesian coordinates and 
then couple the spin and relative momentum in the appropriate way.
For better distinction, we use Greek indices for the spherical representation 
and Latin indices for the Cartesian representation throughout this paper.
The Cartesian form of the strong $D$-wave interaction is taken from 
Refs.~\cite{bedaque2003narrow, brown2014field}
\begin{align}
\label{eq:lagrangianD}
\frac{1}{2} \left( \overset{\leftrightarrow}\nabla_i 
\overset{\leftrightarrow}\nabla_j
+ \overset{\leftrightarrow}\nabla_j \overset{\leftrightarrow}\nabla_i \right)
- \frac{1}{\mathrm{d}-1} \overset{\leftrightarrow}\nabla^2 \delta_{ij} \ ,
\end{align}
where $\mathrm{d}$ denotes the space-time dimension. This interaction yields 9 
components, but a straightforward check shows
that only 5 of them are linearly independent. Thus, the $D$-wave part of the 
Lagrangian~\eqref{eq:lagrangianD} is
Galilei invariant and contains the correct number of degrees of freedom.

The relation between spherical and cartesian coordinates is
given by
\begin{align}
  \label{eq:CStrafo}
  r_{\pm 1}= \mp (x_1\pm ix_2)/\sqrt{2}\,, \quad r_0 = x_3\,,
\end{align}
and similar relations apply to other quantities. For convenience, we will
always use the Cartesian representation, but we will switch to a spherical basis 
if a coupling to definite angular momentum is required.

%%%%%%%%%%%%%%%%%%%%%%%%%%%%%%%%%%%%%%%%%%%%%%%%%%%%%%%%%%%%%%%%%%%%%%%%%%%%%%%%
%%%%%%%%%%%%%%%%%%%%%%%%%%%%%%%%
\begin{figure}[t]
\centering
\includegraphics[width=0.7\textwidth]{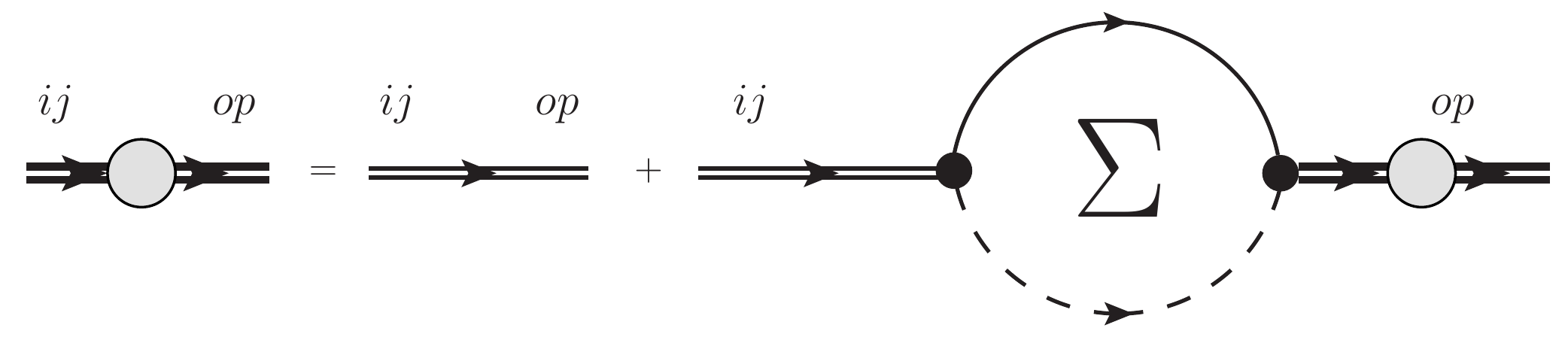}
\includegraphics[width=0.4\textwidth]{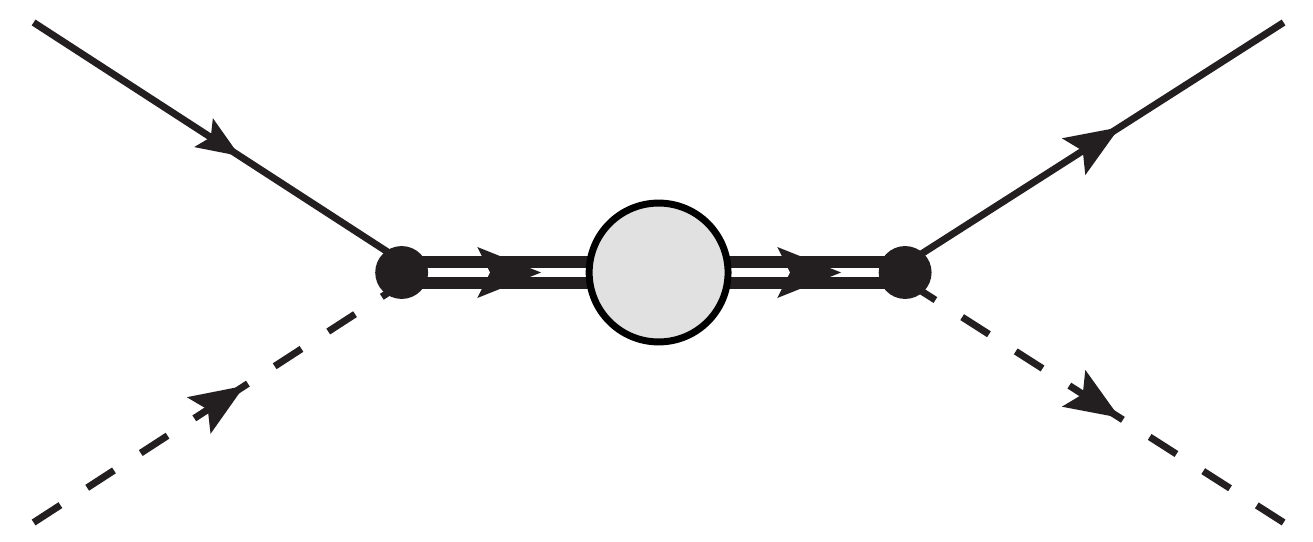}
\caption{The dashed line denotes the core field $c$ and the thin solid line 
the 
neutron. The thin double line represents
the bare dimer propagator and the thick double line with the gray circle is 
the dressed dimer propagator. The top panel shows the diagrammatic 
representation of the Dyson equation for the dressed dimer propagator and the 
bottom panel the neutron-core scattering amplitude with the dressed dimer 
propagator.}
\label{fig:dimer-propagator}
\end{figure}
%%%%%%%%%%%%%%%%%%%%%%%%%%%%%%%%%%%%%%%%%%%%%%%%%%%%%%%%%%%%%%%%%%%%%%%%%%%%%%%%
%%%%%%%%%%%%%%%%%%%%%%%%%%%%%%%%

\subsection{\textit{D}-wave propagator}

The dressed $d$ propagator and the $D$-wave scattering amplitude are computed 
from summing the bubble diagrams in Fig.~\ref{fig:dimer-propagator}. This 
corresponds to the exact solution of the field theory defined by 
the terms explicitly shown
in Eq.~(\ref{eq:lagrangian}) for the $D$-wave state. 
In the next subsection, we will develop a power counting scheme that
classifies the different contributions to the propagator according to their 
importance at low energies. After this scheme has been 
established, only the terms contributing to the considered order 
will be included.
To make the divergence structure transparent, we 
will use a simple momentum cutoff to regularize the loop integrals.
As before, we calculate the dressed $D$-wave propagator in the Cartesian 
representation and couple the neutron spin and the relative momentum to project 
out the appropriate angular momentum $J$ in the end.
%%%%%%%%%%%%%%%%%%%%%%%%%%%%%%%%%%%%%%%%%%%%%%%%%%%%%%%%%%%%%%%%
%%%%%%%%%%%%%%%% at first with cutoff %%%%%%%%%%%%%%%%%%%%%%%%
%%%%%%%%%%%%%%%%%%%%%%%%%%%%%%%%%%%%%%%%%%%%%%%%%%%%%%%%%%%%%%%%
The Dyson equation for the $D$-wave is illustrated in the top panel of 
Fig.~\ref{fig:dimer-propagator} and yields
\begin{align}
\label{eq:d-wave-tensor}
D_d(p)_{ijop} &= D_d(p) \ \frac{\left(\delta_{io} \delta_{jp} + \delta_{ip} 
\delta_{jo}
- \frac{2}{3} \delta_{ij} \delta_{op}\right)}{2} \ , \\
D_d(p) &= \frac{1}{\Delta_2 + \eta_2 \left[p_0 -\mathbf{p}^2/(2 M_{nc}) \right]
+ c_2 \left[p_0 -\mathbf{p}^2/(2 M_{nc}) \right]^2 - \Sigma_d(p)} \ ,
\end{align}
with the one-loop self-energy
\begin{align}
\Sigma_d(p)_{ijop} =& \Sigma_d(p) \ \frac{\left(\delta_{io} \delta_{jp} + 
\delta_{ip} \delta_{jo}
- \frac{2}{3} \delta_{ij} \delta_{op}\right)}{2} \ , \\
\notag
\Sigma_d(p) =& - \frac{m_R g_2^2}{15 \pi} \left[ \frac{2}{5 \pi} \Lambda^5
+ \frac{2}{3 \pi} (2m_R) \left(p_0 - \frac{p^2}{2 M_{nc}} \right) \Lambda^3
+ \frac{2}{\pi} (2m_R)^2 \left(p_0 - \frac{p^2}{2 M_{nc}} \right)^2 \Lambda 
\right. \\
\label{eq:d_s-en_cutoff}
& \left.+ i (2m_R)^{5/2} \left(p_0-\frac{p^2}{2M_{nc}} \right)^{5/2} \right] \ ,
\end{align}
where $m_R=(m_n M)/(m_n+M)$ denotes the reduced mass of the neutron-core system and 
$\Lambda$
is a momentum cutoff.
In spherical coordinates the Cartesian tensor~\cite{brown2014field}
\begin{align}
\label{eq:d_Cartesian_tensor}
\frac{\left(\delta_{io} \delta_{jp} + \delta_{ip} \delta_{jo} - \frac{2}{3} 
\delta_{ij} \delta_{op}\right)}{2}~,
\end{align}
transforms to
\begin{align}
\label{eq:d_spherical_tensor}
\sum_{\alpha \beta \gamma \delta} \left(\left.1 \alpha \ 1 \beta \right\vert 2 
m_l \right)
\left(\left.1 \gamma \ 1 \delta \right\vert 2 m_l' \right) 
\frac{\left(\delta_{\alpha\gamma} \delta_{\beta\delta}
+ \delta_{\alpha\delta} \delta_{\beta\gamma} + 
\delta_{\alpha -\beta}(-1)^{\beta}\delta_{\gamma -\delta}(-1)^{\delta}\right)}{2} \ = \delta_{m_l, m_l'} 
\ 
,
\end{align}
and eventually, the full angular momentum coupling \eqref{eq:d_tensor_L} 
applied 
in the incoming and outgoing channel yields
\begin{align}
\Sigma_d(p)_{m m'} &= \sum_{m_s m_l m_s' m_l'} \left(\left.\frac{1}{2} m_s \ 2 
m_l \right\vert J m \right)
\left(\left.\frac{1}{2} m_s' \ 2 m_l' \right\vert J m' \right) \delta_{m_l 
m_l'} 
\delta_{m_s m_s'} \Sigma_d(p)
= \delta_{m m'} \Sigma_d(p) \ ,
\end{align}
where $m_s$ $(m_s')$ and $m_l$ $(m_l')$ are the spin projections of the created 
(annihilated) neutron and the projections of the $D$-wave interaction at both vertices of the bubble diagram
in Fig.~\ref{fig:dimer-propagator}, respectively. Moreover, $J$
denotes the total spin with its incoming and outgoing projections $m$ and $m'$, respectively. 
The $D$-wave scattering amplitude in the two-body center-of-mass frame with 
$E=k^2/(2m_R)$ and $k=|\mathbf{p}'|=|\mathbf{p}|$ for on-shell scattering
\begin{align}
\label{eq:dwave_matching}
t_2(\mathbf{p}', \mathbf{p}; E) = g_2^2 \left[ \left(\mathbf{p} \cdot 
\mathbf{p}'\right)^2
- \frac{1}{3} \mathbf{p}^2 \mathbf{p}'^2\right] D_d (E,\mathbf{0}) \ ,
\end{align}
is matched to the effective range expansion
\begin{align}
\label{eq:dwave_matchingERE}
t_2(\mathbf{p}', \mathbf{p}; E) = \frac{15 \pi}{m_R} \frac{ \left(\mathbf{p} 
\cdot \mathbf{p}'\right)^2
- \frac{1}{3} \mathbf{p}^2 \mathbf{p}'^2}{1/a_2 - \frac{1}{2} r_2 k^2 + 
\frac{1}{4} \mathcal{P}_2 k^4 + i k^5} \ ,
\end{align}
and we find the matching relations
\begin{align}
\frac{1}{a_2} = \frac{15 \pi}{m_R g_2^2} \Delta_2 + \frac{2}{5 \pi} \Lambda^5\ 
, 
\quad r_2 =
- \frac{15 \pi}{m_R^2 g_2^2} \eta_2 - \frac{2}{3 \pi} \Lambda^3 \ , \quad 
\mathcal{P}_2 = \frac{15 \pi}{m_R^3 g_2^2} c_2 + \frac{2}{\pi} \Lambda \ ,
\end{align}
which determine the running of the coupling constants $g_2$, $\Delta_2$, and 
$c_2$ with the cutoff $\Lambda$.
Since we get $\Lambda$ dependencies with powers of $5$, $3$, and $1$, the 
effective range parameters $a_2$, $r_2$, and $\mathcal{P}_2$ are required
for renormalization at LO.
This pattern motivates our power counting scheme discussed below.
In particular, we include the 2nd-order kinetic term proportional to 
$c_2$ (cf.~Ref~\cite{beane2001rearranging})
in \eqref{eq:lagrangian} in order to absorb the quintic divergence.
If the values for these ERE parameters are known, they can be used to fix the 
EFT couplings $\Delta_2$, $c_2$ and $g_2$ in our
theory.\footnote{Note that we chose $\eta_2$ to be a sign.}
In the vicinity of the bound state pole, the dressed $d$ propagator can be 
written as
\begin{align}
\notag
D_d(p) &= Z_d \ \frac{1}{p_0-\frac{\mathbf{p}^2}{2 M_{nc}}+B_2} + R_d(p) \ , 
\\[6pt]
\label{eq:zd}
Z_d &= -\frac{15 \pi}{m_R^2 g^2_2} \ \frac{1}{r_2 + \mathcal{P}_2 \gamma_2^2 - 
5 
\gamma_2^3} \ ,
\end{align}
{where $Z_d$ denotes the wave-function renormalization, $B_2=\gamma_2^2/(2m_R)$ 
denotes the binding energy with the binding momentum $\gamma_2 \sim 1/R_{halo}$, and $R_d(p)$ is the
remainder which is regular at the pole. The pole condition gives the relation between 
the effective range parameters $a_2$, $r_2$, $\mathcal{P}_2$ and the binding momentum $\gamma_2$
\begin{align}
\frac{1}{a_2}+\frac{1}{2}r_2\gamma_2^{2}+\frac{1}{4}\mathcal{P}_2\gamma_2^{4}=0 \ .
\end{align}}

\subsection{Power counting}
\label{sec:power_counting}

For the shallow $S$-wave state, we adopt the standard power counting
from pionless EFT~\cite{vanKolck:1997ut,vanKolck:1998bw,kaplan1998new,kaplan1998two}.
This implies the scaling $1/\gamma_0 \sim a_0 \sim R_{halo}$ and $r_0 \sim 
R_{core}$, where
$\gamma_0=(1-\sqrt{1-2r_0/a_0})/r_0$ is the bound/virtual state pole
position, $a_0$ the scattering length, and $r_0$ the effective range.
As a result, $r_0$ contributes at next-to-leading order
(NLO) in the expansion in $R_{core}/R_{halo}$.

Because more effective range parameters are involved, the power counting for 
shallow states in higher partial waves is not unique
and different scenarios are conceivable~\cite{bertulani2002effective, 
bedaque2003narrow}.
We apply the constraint that our scheme should exhibit the minimal number of 
fine tunings in the coupling constants
required to absorb all power law divergences.
This is motivated by the expectation that every additional fine tuning makes a 
scenario less likely to be found in nature,
as discussed by Bedaque et al. in Ref.~\cite{bedaque2003narrow}.
They explicitly consider $P$-waves where both $a_1$ and $r_1$ enter at leading 
order (LO)
and assume the scaling relations $a_1\sim R_{halo}^2 R_{core}$ and $r_1 \sim 
1/R_{core}$, while higher effective range parameters scale with the
appropriate power of $R_{core}$. This scenario requires only one fine-tuned 
combination of coupling constants
in contrast to the alternative scenario proposed in 
Ref.~\cite{bertulani2002effective} which requires two.
In this work, we follow the general arguments of Ref.~\cite{bedaque2003narrow} 
and apply them to the $D$-wave case.

To renormalize all divergences in the $D_d(p)$ propagator, 
Eq.~(\ref{eq:d-wave-tensor}), the effective range parameters $a_2$, $r_2$, and 
$\mathcal{P}_2$
are all required.
In the minimal scenario thus two out of three combinations of coupling 
constants 
need to be fine-tuned, i.e. $a_2 \sim R_{halo}^4 \ R_{core}$ and
$r_2 \sim 1/(R_{halo}^2 \ R_{core})$, while $\mathcal{P}_2 \sim 1/R_{core}$. 
With this scaling, all three terms contribute at the same
order for typical momenta $k\sim 1/R_{halo}$.
Higher effective range parameters scale with $R_{core}$ only and thus are 
suppressed by powers of $R_{core}/R_{halo}$.

This means that the dominant contribution to the $D$-wave bound state,
after resumming all bubble diagrams and appropriate renormalization, comes
from the bare propagator. In particular, the general structure of
the propagator near the bound state pole at $E=-B_d$,
\begin{equation}
D_d(E) = \frac{Z_d}{E+B_d} \quad +
\quad\mbox{regular terms in $E$}\,,
\end{equation}
with $Z_d$ the wave function renormalization constant,
is fully reproduced by the bare $D$-wave propagator. Furthermore,
all imaginary parts, if present, appear in the regular part
of the amplitude. With our assumptions about the scaling of
$a_2$, $r_2$, and $\mathcal{P}_2$, the loop contributions are
suppressed by $R_{core}/R_{halo}$. They can be treated in perturbation
theory and contribute at NLO in the power counting. Thus, the
low-energy
$D$-wave scattering amplitude will satisfy unitarity perturbatively
in the expansion in $R_{core}/R_{halo}$.
This is similar to the treatment of the excited $P$-wave bound state
in Ref.~\cite{hammer2011electric}.

We note that the power counting depends sensitively on the details of the 
considered system and thus has to be verified
\textit{a posteriori} by comparison to experimental information.
Including $S$-waves and switching to the pole momentum $\gamma_0$ instead of 
the scattering length $a_0$, the relevant parameters
in our EFT are $\gamma_0$, $\gamma_2$, $r_2$, and $\mathcal{P}_2$ at LO, and 
the following wave function renormalization constants for the $D$-wave state 
are obtained
\begin{align}
\label{eq:normalization}
Z_d = -\frac{15\pi}{m_R^2 g^2_2} \ \frac{1}{r_2 + \mathcal{P}_2 \gamma_2^2} 
\quad \mbox{(LO)} \quad\mbox{and}\qquad
Z_d = -\frac{15\pi}{m_R^2 g^2_2} \ \frac{1}{r_2 + \mathcal{P}_2 \gamma_2^2} \left[1+\frac{5\gamma_2^3}{r_2 + \mathcal{P}_2 \gamma_2^2}\right] \quad\mbox{(NLO)}\ .
\end{align}
The corresponding constants for the $S$-wave state are given in Appendix 
\ref{app:Swave}.

%%%%%%%%%%%%%%%%%%%%%%%%%%%%%%%%%%%%%%%%%%%%%%%%%%%%%%%%%%%%%%%%
%%%%%%%%%%%%%%%% PDS %%%%%%%%%%%%%%%%%%%%%%%%
%%%%%%%%%%%%%%%%%%%%%%%%%%%%%%%%%%%%%%%%%%%%%%%%%%%%%%%%%%%%%%%%
After we have identified the proper power counting, we switch to dimensional 
regularization
with power divergence subtraction
(PDS) with renormalization scale $\mu$ as our regularization 
scheme~\cite{kaplan1998new, kaplan1998two}.
This simplifies the calculations but still keeps the linear divergence 
associated with $\mathcal{P}_2$.
In PDS, the one-loop self-energy for the $D$-wave state is given by
\begin{align}
\Sigma_d(p) = - \frac{2}{15} \frac{m_R g_2^2}{2 \pi} \ (2m_R)^2 \left(p_0 - 
\frac{p^2}{2 M_{nc}} \right)^2
\left[i \sqrt{2m_R\left(p_0-\frac{p^2}{2M_{nc}} +i\epsilon \right)} - 
\frac{15}{8} \mu \right] \ .
\end{align}
After matching to the effective range expansion, we find
\begin{align}
\label{eq:d_PDS_ERE_matching}
\frac{1}{a_2} = \frac{15 \pi}{m_R g_2^2} \Delta_2 \ , \quad r_2 = - \frac{15 
\pi}{m_R^2 g_2^2} \eta_2 \ ,
\quad \mathcal{P}_2 = \frac{15 \pi}{m_R^3 g_2^2} c_2 + \frac{15}{2} \mu \ .
\end{align}
However, we note that the $\mu$ dependence in the matching condition for $\mathcal{P}_2 
\sim 1/R_{core}$ is subleading for $\mu \sim 1/R_{halo}$. It appears only at NLO 
where it is required to absorb the divergence at this order. 
Our findings in the form factor 
calculation confirm this observation.

As pointed out in the introduction, an EFT for $D$-wave states was previously 
considered in the context
of the reaction $d+t \leftrightarrow n+\alpha$ in Ref.~\cite{brown2014field}, 
where the coupling of the
auxiliary field for the $^5$He resonance to the $\alpha n$ pair with spin $3/2$ 
involves a $D$-wave.
We note that in Ref.~\cite{brown2014field} the minimal subtraction (MS) scheme 
was used, in which all power
law divergences are automatically set to zero and no
explicit renormalization is required for the $D$-wave propagator.
As argued in
Refs.~\cite{vanKolck:1997ut,vanKolck:1998bw,kaplan1998new, kaplan1998two},
the MS
scheme is not well suited for systems with shallow bound states since the 
tracking of power law divergences is important.
If MS is used in the $D$-wave case, the contributions of $r_2$ 
and $\mathcal{P}_2$ appear shifted to higher orders. Using a momentum cutoff
scheme
for the $D$-wave
propagator, it becomes clear that contact interactions corresponding to
$r_2$ and $\mathcal{P}_2$ are also required to absorb all divergences at
leading order. As a consequence, these parameters have to be enhanced by the
halo scale $R_{halo}$.

\subsection{Higher partial waves}
\label{sec:higher_partial_waves}
It is straightforward to extend our power counting arguments to partial waves 
beyond the $D$-wave.
The higher-$l$ interaction terms can be derived from the Cartesian (Buckingham) tensors~\cite{buckingham1959molecular, Briceno:2013lba}, 
which are symmetric and traceless
in every pair of  indices and are
given by
\begin{equation}
M_{i, j, \cdots, t}^{(l)} = \frac{(-1)^l}{l!} r^{2l+1} 
\frac{\partial^l}{\partial x_i \partial x_j \cdots \partial x_t}
\left(\frac{1}{r}\right) \qquad i, j, \cdots , t \in \{1,2,3\} \ 
\end{equation}
with $r=\sqrt{x_1^2+x_2^2+x_3^2}$.
To obtain the specific interaction in momentum space for a given angular 
momentum $l$, $r_j$ is simply replaced by
$i \nabla_j$. In general, this leads to a tensor of rank $l$ with $3^l$ 
components. However, because the tensors
are symmetric and traceless in every pair of indices, only $[(2+l)(1+l)..3]/l! 
- 
l(l-1)/2=2l+1$ components are linearly independent.
Thus, we obtain the correct number of linearly independent components for a 
given partial wave.
However, beyond $P$-waves it becomes beneficial to use spherical representation 
for calculations
depending on the considered observable.

In order to be able to absorb all power law divergences, the first $(l+1)$ 
effective range parameters are needed at LO
for the $l$-th partial wave~\cite{bertulani2002effective}.
As discussed in the previous subsection, one of these parameters can be assumed 
to scale only with $R_{core}$ if $l \geq 1$.
Thus, we need $l$ fine-tuned parameters for the $l$-th partial wave if we want 
to renormalize all power law
divergences assuming the minimal fine tuning scenario.
For arbitrary $l$, this leads to the following power counting scheme
\begin{align}
\label{eq:powerCounting1}
a_l &= \begin{cases}
R_{halo}, & l=0 \\
R_{halo}^{2l} \ R_{core}, & l>0
\end{cases}\\[10pt]
r_l &= \begin{cases}
R_{core}, & l=0 \\
1/\left(R_{halo}^{2l-2} \ R_{core}\right), & l>0
\end{cases}\\[10pt]
\label{eq:powerCounting2}
\mathcal{P}_l &= \begin{cases}
R_{core}^{3-2l}, & l\leq1 \\
1/\left(R_{halo}^{2l-4} \ R_{core}\right), & l>1
\end{cases}\\[8pt]\notag
& \qquad \qquad \vdots \ ,
\end{align}
where the $l$-th and higher effective range parameters in each partial wave 
scale with appropriate powers
of $R_{core}$. Accordingly, our power counting scenario agrees with 
Ref.~\cite{bedaque2003narrow} up to $P$-waves
but differs beyond that because the higher effective range parameters are 
counted
differently. The condition that all power law divergences in the bubble diagram 
can be absorbed is relaxed and
only $a_l$ and $r_l$ contribute at LO for arbitrary $l>0$. As a consequence, 
only 
one fine tuning is required.
If this counting is universally realized in nature, one would expect an 
approximately equal number of shallow states
in low and high $l$-waves. Since experimental observation of shallow states in light nuclei
is predominated by lower $l$-waves, we expect our counting to be more realistic. Later
in our calculations for $^{15}$C we compare both power countings and reveal that our scenario
is more compatible with data in this case.

\section{Electric observables}
\label{sec:EM_Observables}
In this section, we use the Lagrangian \eqref{eq:lagrangian} with minimal 
substitution plus the local, gauge invariant operators 
to compute the $D$-wave form factors and the E2 transition from the $S$- to the 
$D$-wave states. Eventually, we apply our results to $^{15}$C to predict 
several electric observables.

\subsection{Electric interactions}
\label{sec:EM_interactions}

Electric interactions are included via minimal substitution in the 
Lagrangian
\begin{align}
\partial_\mu \rightarrow D_\mu = \partial_\mu + i e\hat{Q} A_\mu \ ,
\end{align}
where the charge operator $\hat{Q}$ acting on the ${}^{14}$C core yields 
$\hat{Q} c = 6 c$. Additionally to the electric interactions resulting 
from the application of the minimal substitution in the Lagrangian, we have to 
consider further local gauge invariant operators involving the electric field 
$\mathbf{E}$ and the fields $c$, $n$, $\sigma$ and $d$. Depending on the 
observable and respective partial wave, they contribute at different orders of 
our EFT. The local operators with one power of the photon field, relevant in 
our calculation of electric form factors and the B(E2) transition strength, are
 \begin{align}
 \notag
 \mathcal{L}_{E} =& - L_{C01}^{(d)} \ d_{m}^\dagger \left(\nabla^2 A_0 - 
\partial_0 (\boldsymbol{\nabla} \cdot \mathbf{A})\right) d_m\\[6pt]\notag
 & - L_{C02}^{(d)} \ d_{m'}^\dagger \left(\left. \frac{1}{2} m_s \ 2 m_{l'} 
\right\vert J m' \right) \left(\left. \frac{1}{2} m_s \ 2 m_l \right\vert J m 
\right) \left(\left. 1 \alpha \ 1 \beta \right\vert 2 m_{l'} \right) \left(\left. 
1 \gamma \ 1 \delta \right\vert 2 m_l \right) \\[4pt] \notag
 & \quad \left[ \left(\nabla_\alpha \nabla_\gamma \delta_{\beta\delta} A_0 
\right) - \left( \partial_0 \frac{\left(\nabla_\alpha A_\gamma +\nabla_\gamma 
A_\alpha\right)}{2} \delta_{\beta\delta} \right) \right] d_m\\[6pt]
 \label{eq:localOperators}
 & -L_{E2}^{(sd)} \sigma_m d^\dagger_{m'} \left(\left. \frac{1}{2} m \ J m' 
 \right\vert 2 m_l \right) \left(\left. 1 \alpha \ 1 \beta \right\vert 2 m_l 
\right) \left[ \nabla_\alpha \nabla_\beta A_0 - \partial_0 \frac{\left( 
\nabla_\alpha A_\beta + \nabla_\beta A_\alpha \right)}{2} \right],
 \end{align}
where repeated spin indices are summed over.

These additional operators are necessary in order to renormalize our
results in the electric sector.
This means that up to a certain order within our power counting, ultraviolet
divergences can only be removed through interactions as in Eq.
\eqref{eq:localOperators}.
In particular, the interaction terms proportional to $L_{C01}^{(d)}$ and
$L_{C02}^{(d)}$ are required in order to remove the divergences occurring in
the loop diagram in Fig.~\ref{fig:formfactors}. Since this loop diagram is a
NLO contribution, the corresponding interaction terms are entering first at
NLO.
This procedure allows us to determine the highest possible order within our
power counting scheme that the interactions in Eq. \eqref{eq:localOperators}
have to enter in order to eliminate divergences.

\subsection{E2 transition}

\begin{figure}[t]
\centering
\includegraphics[width=0.65\textwidth]{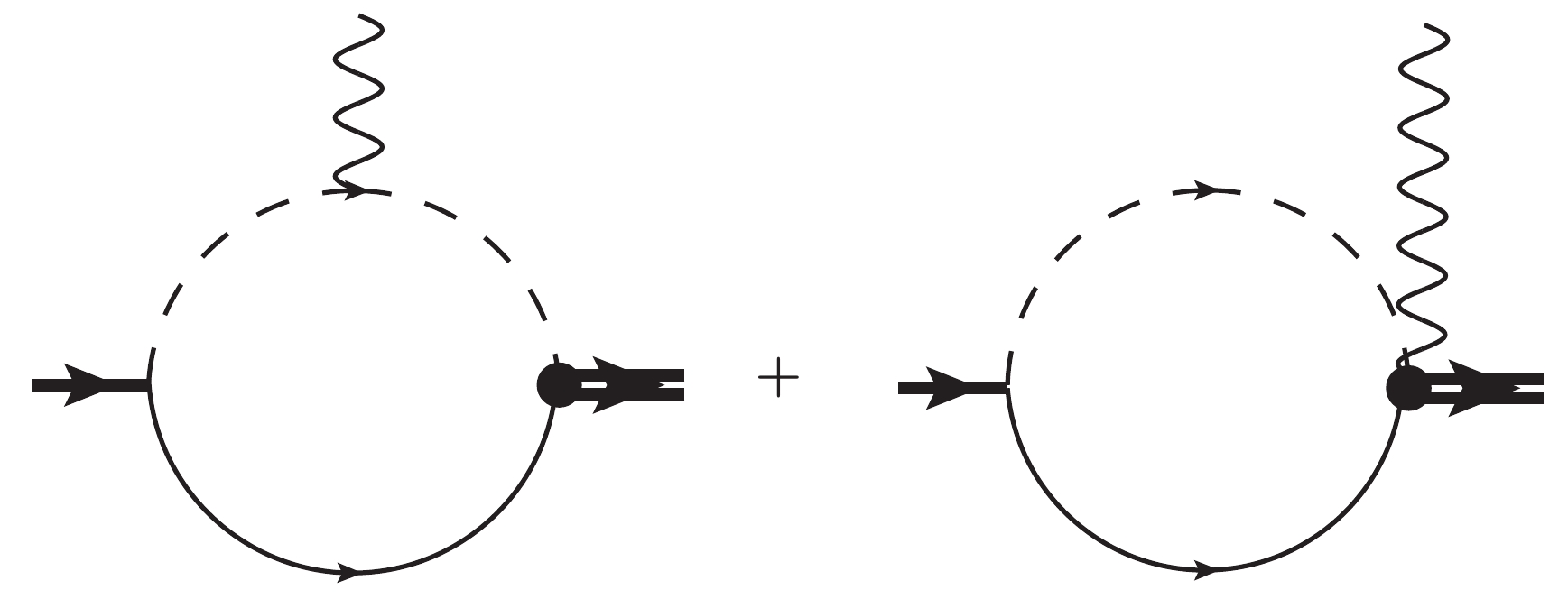}
\caption{The diagrams contributing to the irreducible vertex that determines 
the 
$S$-to-$D$ state transition in Halo EFT. The thick double line denotes the 
dressed $D$-wave propagator and the thick single line the dressed $S$-wave 
propagator.}
\label{fig:E2trans}
\end{figure}

The diagrams contributing to the irreducible vertex for
the E2 transition from the $S$- to the $D$-wave state
at LO are shown in Fig.~\ref{fig:E2trans}. At higher order,
the next contribution would be the counterterm $L_{E2}^{(sd)}$
from Eq.~(\ref{eq:localOperators}) which has to be fixed by
experimental input. The interaction term proportional to
$L_{E2}^{(sd)}$ is not required to cancel any divergence because the LO
contributions to the B(E2) transition depicted in Fig.~\ref{fig:E2trans}
are finite. Therefore, our minimal principle of including the counterterms
when they are needed for renormalization cannot be used to determine its
exact order beyond LO.
Thus, we restrict ourselves to LO for the reduced E2 transition strength.
This allows us to make predictions for other electric
observables as discussed below.

The photon in Fig.~\ref{fig:E2trans} has a four momentum of $k=(\omega, \mathbf{k})$ 
and its polarization index is denoted by $\nu$.
The computation of the relevant diagrams yields a vertex function $\Gamma_{m' 
m_s \nu}$, where $m'$ is the total angular momentum projection of the $D$-wave 
state and $m_s$ denotes the spin projection of the $S$-wave state.
Since the neutron spin is unaffected by this transition, we calculate the 
vertex 
function with respect to the specific components of the $D$-wave interaction
\begin{align}
\label{eq:E2vertex}
\Gamma_{m' m_s \nu} = \sum_{m_l} \left(\left. \frac{1}{2} m_s \ 2 m_l 
\right\vert J m' \right) \sum_{\alpha \beta} \left(\left. 1 \alpha \ 1 \beta 
\right\vert 2 m_l \right) \tilde{\Gamma}_{\alpha\beta\nu} \ ,
\end{align}
where $J$ denotes the total spin of the $D$-wave state.
In the case of $m_s = m'= \pm 1/2$, only $m_l = 0$ contributes to the sum in 
Eq.~\eqref{eq:E2vertex} and we get for $J = 5/2$
\begin{align}
\label{eq:GammaVertex1}
\Gamma_{+\frac{1}{2}\,+\frac{1}{2}\,\nu} = \Gamma_{-\frac{1}{2}\,-\frac{1}{2}\,\nu} = \sqrt{\frac{2}{5}} \tilde{\Gamma}_{00\nu}\ + 
\sqrt{\frac{1}{10}} \tilde{\Gamma}_{1-1\nu} + \sqrt{\frac{1}{10}} 
\tilde{\Gamma}_{-11\nu}\ .
\end{align}
We 
calculate the irreducible vertex in Coulomb gauge so that we have $\mathbf{k} 
\cdot \mathbf{\epsilon} = 0$ for real photons.
In order to isolate the electric contribution to the
irreducible vertex in a simple way,
we choose $\mathbf{k} \cdot \mathbf{p} = 0$,
where $\mathbf{p}$ denotes the incoming momentum of the $S$-wave
state.
Taking gauge invariance and symmetry properties into account, the 
space-space components of the vertex function in Cartesian coordinates can be 
written as
\begin{align}
\label{eq:gammaE-gammaM}
&\tilde{\Gamma}_{ijk} = \Gamma_E \ \frac{1}{2} \left( k_j \delta_{ik} + k_i 
\delta_{jk} \right) \ + \ \Gamma_M \ p_k \left( k_i k_j \ - \frac{1}{3} 
\delta_{ij} k^2 \right).
\end{align}
Choosing the photon to be traveling in the $x_3$ direction 
only $\Gamma_E$ contributes to $\tilde{\Gamma}_{33 3}$, and we obtain
\begin{align}
\label{eq:GammaVertex2}
\tilde{\Gamma}_{33 3} = \Gamma_E \omega \ ,
\end{align}
with $|\mathbf{k}| = k_3 = \omega$.
By comparing the definitions for the transition rate depending on B(E2) and the 
transition rate as a function of the irreducible vertex 
$\Gamma_E$~\cite{greiner1996nuclear}, we get the following relation
\begin{align}
\notag
\text{B(E2: $1/2^+ \rightarrow 5/2^+$)} &= \frac{15}{\pi} \ 
\left(\frac{\Gamma_{+\frac{1}{2}\,+\frac{1}{2}\,0}}{\omega^2}\right)^2 \ .
\end{align}
Evaluating this using  Eqs.~(\ref{eq:GammaVertex1}, \ref{eq:GammaVertex2}), it follows
that $\Gamma_{+\frac{1}{2}\,+\frac{1}{2}\,0}=\sqrt{2/5}\;\tilde{\Gamma}_{33 3}$
and we obtain
\begin{align}
\notag
\text{B(E2: $1/2^+ \rightarrow 5/2^+$)} &= \ \frac{6}{\pi} \ 
\left(\frac{\bar{\Gamma}_E}{\omega}\right)^2 \ ,
\end{align}
with the renormalized, irreducible vertex $\bar{\Gamma}_E = \sqrt{Z_\sigma Z_d} 
\, 
\Gamma_E$. At LO, $Z_\sigma$ and $Z_d$ are given in 
Eqs.~\eqref{eq:zs} and \eqref{eq:normalization}, respectively.
Using the result of our calculation of $\Gamma_E$ for diagrams (a), we find at 
LO
\begin{align}
\label{eq:E2_LO}
\text{B(E2: $1/2^+ \to 5/2^+$)} &= \frac{4}{5 \pi} \frac{Z_\text{eff}^2 e^2 
\gamma_0}{-r_2 -\mathcal{P}_2 \gamma_2^2} \ \left[ \frac{3\gamma_0^2 + 
9\gamma_0\gamma_2 + 8\gamma_2^2}{(\gamma_0 + \gamma_2)^3} \right]^2 ,
\end{align}
where $\gamma_2$, $r_2$, and $\mathcal{P}_2$ are the parameters of the $5/2^+$ 
state and
the effective charge is $Z_\text{eff} = (m_n/M_{nc})^2 Q_c$. In general, the 
effective charge
for arbitrary multipolarity $\lambda$ is
given by $Z_\text{eff}^{(\lambda)} = Z_n \left(\frac{M}{M_{nc}}\right)^\lambda 
+ 
Z_c \left(- \frac{m_n}{M_{nc}}\right)^\lambda$
~\cite{typel2005electromagnetic}. In Halo EFT it comes automatically out of the 
calculation.

The same result for $\bar{\Gamma}_E$ can be obtained using current conservation,
\begin{align}
\omega \tilde{\Gamma}_{ij0} = k_k \tilde{\Gamma}_{ijk} \ ,
\end{align}
if we calculate the space-time components of the vertex function 
$\tilde{\Gamma}$. In contrast to $\tilde{\Gamma}_{ijk}$, we have to consider 
only the left diagram in Fig.~\ref{fig:E2trans} for $\tilde{\Gamma}_{ij0}$ at 
LO.

The calculation of the transition to the $3/2^+$ state can be carried out in 
the same way. The only difference is a relative factor of $2/3$ for B(E2) 
because of the different Clebsch-Gordan coefficient in Eq. 
\eqref{eq:GammaVertex1}
\begin{align}
\label{eq:E2_LO_3/2}
\text{B(E2: $1/2^+ \to 3/2^+$)} &= \frac{8}{15 \pi} \frac{Z_\text{eff}^2 e^2 
\gamma_0}{-r_2 -\mathcal{P}_2 \gamma_2^2} \ \left[ \frac{3\gamma_0^2 + 
9\gamma_0\gamma_2 + 8\gamma_2^2}{(\gamma_0 + \gamma_2)^3} \right]^2 ,
\end{align}
where $\gamma_2$, $r_2$, and $\mathcal{P}_2$ are now the parameters of the 
$3/2^+$ state.

\subsection{Form factors}
\label{sec:formfactors}
\begin{figure}[t]
\centering
\includegraphics[width=0.5\textwidth]{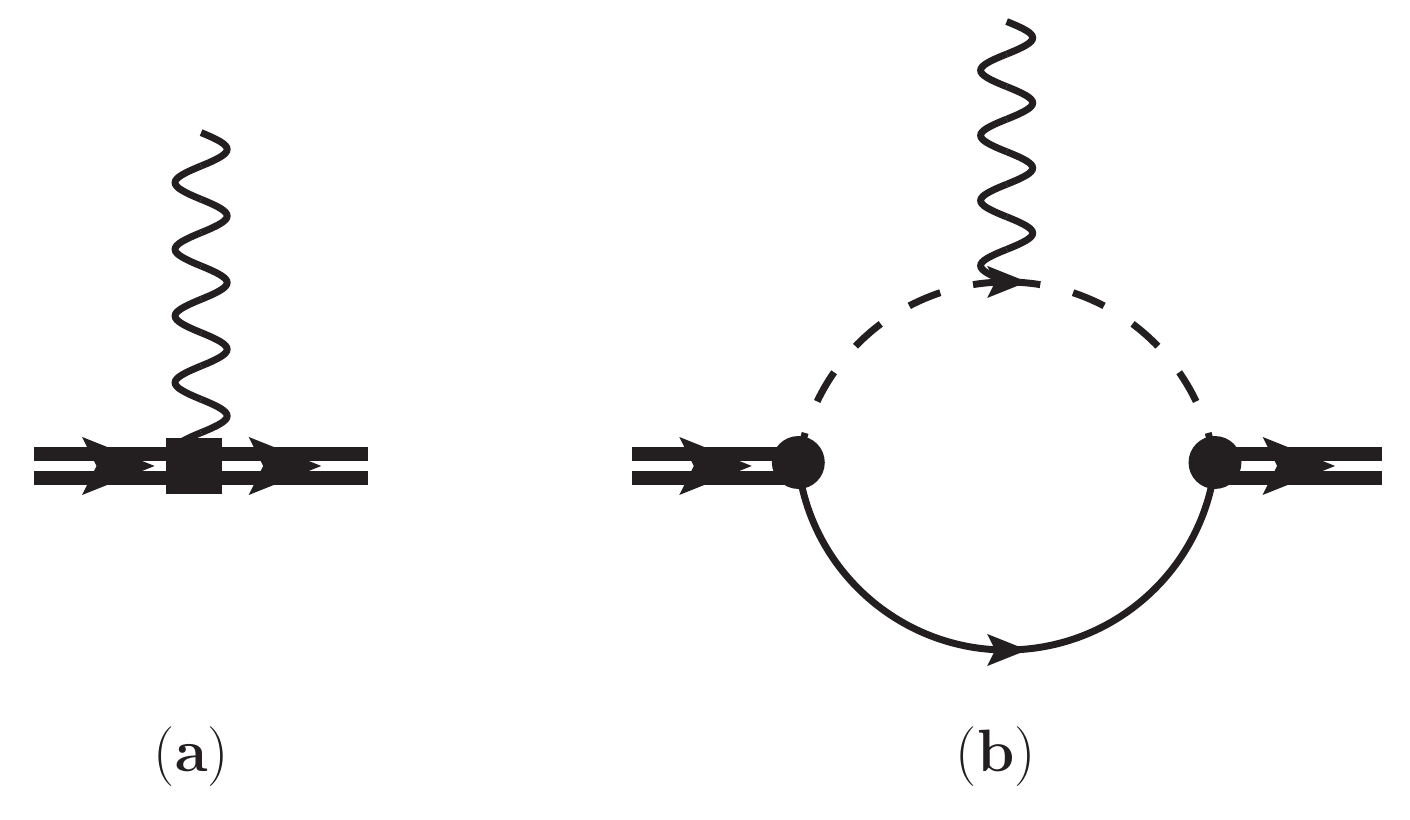}
\caption{The topologies contributing to the irreducible vertex for an $A_0$ 
photon coupling to the ${}^{14}$C-neutron $D$-wave bound state up to NLO.
Diagram (a) contains three different direct couplings. Two arise from minimal 
substitution in the bare propagator proportional to $r_2$, $\mathcal{P}_2$ and 
contribute at LO, while diagram (b) emerges from minimal 
substitution in the core propagator and contributes at NLO. The 
local gauge invariant operator $\sim L_{C01/2}^{(d)}$, required for the 
renormalization of diagram (b), is also represented by diagram (a) and 
contributes at NLO. The thick double line denotes the dressed $D$-wave 
propagator.}
\label{fig:formfactors}
\end{figure}
The result for the electric form factor of an $S$-wave halo state is discussed 
in Ref.~\cite{hammer2011electric} for ${}^{11}$Be and in 
Ref.~\cite{fernando2015electromagnetic} for ${}^{15}$C.
The experimental result for the rms charge radius of $^{14}$C is
$\braket{r_E^2}^{1/2}_{{}^{14}\text{C}} = 2.5025 (87)$ fm~\cite{angeli2013table}
and the Halo EFT result for the  $\frac{1}{2}^+$ $S$-wave ground state is
$\braket{r_E^2}_{{}^{15}\text{C}}^{(\sigma)}\approx 0.11 \
\text{fm}^2$~\cite{fernando2015electromagnetic}, but the authors do not quote an
error for this number.
In principle, both values can be combined to obtain a prediction for the
full charge radius of the $^{15}$C ground state.

Here, we focus on the 
form factors of the $D$-wave state in ${}^{15}$C.
The $D$-wave form factors can be extracted from the irreducible vertex for $A_0 
d d$ interactions. The corresponding contributions are shown in 
Fig.~\ref{fig:formfactors} up to NLO. The first diagram represents three 
different direct couplings of the photon to the $D$-wave propagator. Two 
couplings emerge from the minimal substitution in the bare propagator 
proportional 
to $r_2$, $\mathcal{P}_2$ and contribute at LO. The last one is a
term $\sim L_{C01/2}^{(d)}$ which comes out of Eq.~\eqref{eq:localOperators} 
and is required for the renormalization of the loop divergences of diagram~(b) 
and therefore contributes at NLO. The second diagram arises from minimal 
substitution in the core propagator and contributes at NLO.
The computation is carried out in the Breit frame, $q = (0,\mathbf{q})$, and 
the irreducible vertex for the $A_0$ photon coupling to the $D$-wave state in 
Cartesian coordinates yields
\begin{align}
\Big \langle ij \left| J^0 \right| op \Big \rangle =& - i e Q_c \left[ 
G_E(|\mathbf{q}|) \ E_{ij,op} + \frac{1}{2 M_{nc}^2} G_Q(|\mathbf{q}|) \ 
Q_{ij,op} + \frac{1}{4 M_{nc}^4} G_H(|\mathbf{q}|) \ H_{ij,op} \right] \ ,
\end{align}
with the three-momentum of the virtual photon 
$\mathbf{q}=\mathbf{p'}-\mathbf{p}$ and three different $D$-wave tensors for 
each form factor $E_{ij,op} \sim q^0$, $Q_{ij,op} \sim q^2$ and $H_{ij,op} \sim 
q^4$. Note that we take out an overall factor of the elementary charge $e$ 
from all form factors. As a consequence our definition of the 
quadrupole and hexadecapole moments does not contain a factor $e$.
Evidently, the hexadecapole form factor is only observable for the $5/2^+$ 
$D$-wave state and unobservable for the $3/2^+$ state. This can be 
straightforwardly proven by considering the respective Clebsch-Gordan 
coefficients to couple the spin and angular momentum to total $J$ for the two 
$D$-wave states in combination with $H_{ij,op}$ in spherical coordinates.
For reasons of simplicity, the calculation is carried out in Cartesian 
coordinates and the resulting Cartesian tensors are given below
\begin{align}
\tilde{Q}_{ij,op} &= \frac{1}{4} \left(q_j q_p \delta_{io} + q_j q_o 
\delta_{ip} 
+ q_i q_p \delta_{jo} + q_i q_o \delta_{jp} - \frac{4}{3} q_i q_j \delta_{op} - 
\frac{4}{3} q_o q_p \delta_{ij} + \frac{4}{9} q^2 \delta_{ij} \delta_{op} 
\right) \ ,\\
\tilde{H}_{ij,op} &= \left( q_i q_j q_o q_p - \frac{1}{3} q^2 q_i q_j 
\delta_{op} - \frac{1}{3} q^2 q_o q_p \delta_{ij} + \frac{1}{9} q^4 \delta_{ij} 
\delta_{op} \right) \ ,
\end{align}
\begin{align}
E_{ij,op} &= \frac{\left(\delta_{io} \delta_{jp} + \delta_{ip} \delta_{jo} - 
\frac{2}{3} \delta_{ij} \delta_{op} \right)}{2} \ ,\\[6pt]
Q_{ij,op} &= \frac{3}{5} \tilde{Q}_{ij,op} - \frac{1}{5} q^2 E_{ij,op} \ 
,\\[6pt]
H_{ij,op} &= \frac{3}{2} \tilde{H}_{ij,op} - \frac{30}{35} q^2 
\tilde{Q}_{ij,op} 
+ \frac{3}{35} q^4 E_{ij,op} \ .
\end{align}
The Cartesian tensors $E_{ij,op}$, $Q_{ij,op}$ and $H_{ij,op}$
fulfill the following constraints
\begin{align}
&E_{ij,op} \ E_{ij,op} = 5, \qquad \delta_{ij} \ E_{ij,op} = \delta_{op} \ 
E_{ij,op} = 0 \ ,\\[6pt]
&E_{ij,op} \ Q_{ij,op} = 0, \qquad \delta_{ij} \ Q_{ij,op} = \delta_{op} \ 
Q_{ij,op} = 0 \ ,\\[6pt]
&E_{ij,op} \ H_{ij,op} = 0, \qquad Q_{ij,op} \ H_{ij,op} = 0, \qquad 
\delta_{ij} 
\ H_{ij,op} = \delta_{op} \ H_{ij,op} = 0 \ .
\end{align}
The neutron spin is unaffected by the charge operator up to the order 
considered here.

At LO, only the direct coupling from the minimal substitution in the bare 
$D$-wave propagator proportional to $r_2$ and $\mathcal{P}_2$, 
depicted in Fig.~\ref{fig:formfactors} (a), contribute. This reproduces the 
correct normalization condition of the electric form factor of $G_E(0) = 1$,
but the form factor is just a constant. Therefore, there is no real prediction
beyond charge conservation at LO.

At NLO, diagram (b) in Fig.~\ref{fig:formfactors} also contributes, and the 
counterterm is required for the renormalization of the loop divergences 
stemming from diagram (b).
We then obtain for the electric $G_E(|\mathbf{q}|)$, quadrupole 
$G_Q(|\mathbf{q}|)$ and hexadecapole $G_H(|\mathbf{q}|)$ form factors
\begin{align}
\notag
G_E(|\mathbf{q}|) =& \ \frac{1}{r_2 + \mathcal{P}_2 \gamma_2^2} 
\left[\left(\tilde{L}_{C01}^{(d)} + \frac{4}{3} \tilde{L}_{C02}^{(d)}\right) 
|\mathbf{q}|^2 - \frac{21 \gamma _2 f^2 |\mathbf{q}|^2}{16} + \frac{\gamma 
_2^3}{4} +\gamma _2^2 \mathcal{P}_2 + r_2 \right.\nonumber\\[4pt]
\label{eq:electricFormfactor}
&\left. - \arctan\left(\frac{f |\mathbf{q}|}{2 \gamma _2}\right) \left(\frac{21 
f^3 |\mathbf{q}|^3}{32} + \frac{13 \gamma _2^2 f |\mathbf{q}|}{4} + 
\frac{\gamma 
_2^4}{2 f |\mathbf{q}|} \right) \right] ,\\
G_Q(|\mathbf{q}|) =& \ \frac{2 M_{nc}^2}{r_2 + \mathcal{P}_2 \gamma_2^2} \left[ \frac{20}{3} \tilde{L}_{C02}^{(d)} - \frac{75 \gamma _2 
f^2}{14} - \frac{25 \gamma _2^3}{7 |\mathbf{q}|^2} \right.\nonumber\\[4pt]
\label{eq:quadrupoleFormFactor}
&\left. - \arctan\left(\frac{f |\mathbf{q}|}{2 \gamma _2}\right) \left( 
\frac{75 
f^3 |\mathbf{q}|}{28} +\frac{75 \gamma _2^2 f}{14 |\mathbf{q}|} -\frac{50 
\gamma 
_2^4}{7 f |\mathbf{q}|^3} \right)\right] ,\\
G_H(|\mathbf{q}|) =& \ \frac{2}{3} \ \frac{4 M_{nc}^4}{r_2 + \mathcal{P}_2 
\gamma_2^2} \left[ - \frac{45 \gamma_2 f^2}{64 |\mathbf{q}|^2} 
+\frac{45 \gamma _2^3}{16 |\mathbf{q}|^4}\right.\nonumber\\[4pt]
\label{eq:hexadecapoleFormFactor}
&\left. - \arctan\left(\frac{f |\mathbf{q}|}{2 \gamma _2}\right) \left( 
\frac{45 
f^3}{128 |\mathbf{q}|} -\frac{15 \gamma _2^2 f}{16 |\mathbf{q}|^3} +\frac{45 
\gamma _2^4}{8 f |\mathbf{q}|^5} \right) \right] ,
\end{align}
with $f = m_R/M$ while $\tilde{L}_{C01/2}^{(d)}$ represents the local gauge 
invariant operators from Eq.~\eqref{eq:localOperators}.
These operators have a finite piece $L_{C01/2}^{(d)\text{ fin}}$ as 
well as a $\mu$-dependent part that cancels the renormalization scale
dependence from the loop contribution.
For a better readability, we have absorbed some prefactors in the
definition of the counterterms and defined the low-energy
constants
\begin{align}
\tilde{L}_{C01}^{(d)} &= \frac{15 \pi}{e Q_c g_2^2 m_R^2} L_{C01}^{(d)\text{ 
fin}} \ ,\\[8pt]
\tilde{L}_{C02}^{(d)} &= \frac{15 \pi}{e Q_c g_2^2 m_R^2} L_{C02}^{(d)\text{ 
fin}} \ ,
\end{align}
which are used in Eqs.~(\ref{eq:electricFormfactor},
\ref{eq:quadrupoleFormFactor}, \ref{eq:hexadecapoleFormFactor}).
The remaining divergence emerging from the loop diagram in 
Fig.~\ref{fig:formfactors} is absorbed by $\mathcal{P}_2$ from the direct 
photon coupling in diagram (a).
After the expansion of Eq. \eqref{eq:electricFormfactor}
\begin{align}
G_E(|\mathbf{q}|) \approx 1 - \frac{1}{6} \braket{r_E^2} |\mathbf{q}|^2 + 
\ldots 
\ ,
\end{align}
we obtain $G_E(0) = 1$ and an electric radius
\begin{align}
\label{eq:r_E_LO}
\braket{r_E^2}^{(d)} \ = \ -\frac{12 \tilde{L}_{C01}^{(d)} + 16 \tilde{L}_{C02}^{(d)} - 35 \gamma _2 
f^2}{2 
\left(r_2 + \mathcal{P}_2 \gamma_2^2\right)} \ ,
\end{align}
such that the electric radius is not a prediction.

We expand Eq. \eqref{eq:quadrupoleFormFactor} and 
\eqref{eq:hexadecapoleFormFactor}
\begin{align}
\label{eq:r_Q_LO}
\frac{1}{ M_{nc}^2} G_Q(|\mathbf{q}|) &\approx \mu_Q \left( 1 - \frac{1}{6} 
\braket{r_Q^2} |\mathbf{q}|^2 + \ldots \right) ,\\
\frac{1}{ M_{nc}^4} G_H(|\mathbf{q}|) &\approx \mu_H \left( 1 - \frac{1}{6} 
\braket{r_H^2} |\mathbf{q}|^2 + \ldots \right) ,
\end{align}
and find the respective multipole moments
\begin{align}
\label{eq:NLO_muQ}
\mu_Q^{(d)} &= \frac{10 \left(4 \tilde{L}_{C02}^{(d)} -5 \gamma _2 
f^2\right)}{3 
\left(r_2 + \mathcal{P}_2 \gamma_2^2\right)}~,\\
\label{eq:muH}
\mu_H^{(d)} &= -\frac{2 f^4}{3 \gamma_2 \left(r_2 + \mathcal{P}_2 
\gamma_2^2\right)} \ ,
\end{align}
where we find the hexadecapole moment $\mu_H^{(d)}$ as a prediction.
The electric radius $\braket{r_E^2}^{(d)}$ and the
quadrupole moment $\mu_Q^{(d)}$ are not predicted. They are used to fix the
counterterms $L_{C02}^{(d)}$ and $L_{C01}^{(d)}$.
The quadrupole and hexadecapole radii yield
\begin{align}
\label{eq:rQ}
\braket{r_Q^2}^{(d)} &= \frac{27 f^4}{7 \gamma_2 \left(4 \tilde{L}_{C02}^{(d)} -5 \gamma _2 f^2\right)} \ ,\\
\braket{r_H^2}^{(d)} &= \frac{9 f^2}{14 \gamma_2^2} \ ,
\end{align}
where the hexadecupole radius is predicted by Halo EFT and the quadrupole radius depends on the counterterm $L_{C02}^{(d)}$, fixed by the quadrupole moment. Thus, we can predict the quadrupole radius if the quadrupole moment is known.

Finally, we can reinsert the matching conditions,
Eqs.~(\ref{eq:r_E_LO}, \ref{eq:NLO_muQ}),
into the results for the electric and
quadrupole form factors, Eqs.~(\ref{eq:electricFormfactor},
\ref{eq:quadrupoleFormFactor}), in order to get expressions in terms of
observables only.

\subsection{Correlations between electric observables}
\label{sec:Correlations between EM observables}

Up to this point, all results are universal and not specific for $^{15}$C.
In this section, we explore universal relations between different
observables for shallow $D$-wave bound states predicted by Halo EFT. Moreover,
we combine our Halo EFT results with data and
\textit{ab initio} results from the IT-NCSM~\cite{roth2009importance} to predict
electric properties of $^{15}$C. In a second step, the correlations obtained
in Halo EFT are compared to the E2 correlation based on the rotational
model by Bohr and Mottelson~\cite{bohr1975nuclear}.

We note that the quantification of theory 
uncertainties is important in any application of EFT to actual systems. 
In our discussion below, we will estimate the theory uncertainties 
from the size of the expansion parameter $R_{core}/R_{halo}$. 
More sophisticated estimates can be obtained from Bayesian statistics
\cite{Furnstahl:2015rha,Wesolowski:2015fqa,Griesshammer:2015ahu}, 
but such an analysis is beyond the scope of this manuscript.

To make predictions, we use the experimental transition strength
$\text{B(E2)} = 0.44 (1)$~W.u.~\cite{ajzenberg1991energy} from the $5/2^+ 
\rightarrow 1/2^+$ transition in $^{15}$C to determine the denominator of the 
$D$-wave renormalization constant at LO,
i.e. the combination $r_2+\mathcal{P}_2\gamma_2^2$.
Converting to physical units, we obtain
the strength $\text{B(E2)} = 2.90 (7) \ e^2 \text{fm}^{4}$ for the transition 
$1/2^+ \rightarrow 5/2^+$. The experimental values of the binding momenta are 
$\gamma_0 = 0.235$ fm$^{-1}$ and $\gamma_2 = 0.147$ fm$^{-1}$ 
\cite{ajzenberg1991energy}.
Moreover, Fernando et al.~\cite{fernando2015electromagnetic} argued that it is 
more appropriate to count $r_0 \sim \ R_{halo}$ for the specific case of 
${}^{15}$C and thus we keep this contribution at LO
in our application to $^{15}$C. The extracted value for
$r_0 = 2.67 \ \text{fm}$~\cite{fernando2015electromagnetic} results from a
fit to one-neutron capture data
${}^{14}$C$(n,\gamma){}^{15}$C~\cite{nakamura2009neutron}. With these data, we 
are able to determine
the numerical value for $Z_d m_R^2 g_2^2 \sim 1/\left(r_2 + \mathcal{P}_2 
\gamma_2^2\right) = -181(4) \ \text{fm}^{3}$.

Using our results from the previous sections, we obtain $\tilde{L}_{C01}^{(d)} 
+\frac{4}{3}\tilde{L}_{C02}^{(d)}= \braket{r_E^2}^{(d)} / 1088(25) \ 
\text{fm}^{-1}$, $\tilde{L}_{C02}^{(d)} = - \mu_Q^{(d)} / 2418(55)$ 
$\text{fm}^{-1}$
for the finite piece of the counterterms.
For the hexadecapole moment and radius, we obtain the following predictions
\begin{align}
\mu_H^{(d)} = 1.68(4)(50) \times 10^{-2}\ \text{fm}^{4}\,, \ \mbox{ and } \
\braket{r_H^2}^{(d)} = 0.135(3)(40) \ \text{fm}^{2}\,,
\end{align}
where the first uncertainty is due to the experimental input and the second one
is a theory uncertainty from higher order corrections of order $R_{core}/R_{halo}
\approx 0.3$ (see below).

{Comparing our findings with Ref.~\cite{hammer2011electric} we find, as a 
general 
rule, that the highest multipole form factor is always independent of 
additional 
parameters from short-range counterterms. 
Moreover, we can always find a smooth correlation between the highest radius and the neutron separation energy $S_n$
\begin{align}
 \braket{r_H^2}^{(d)} = \frac{9}{28} \frac{f^2}{m_R S_n^{(d)}}~.
\end{align}
In the $S$- and $P$-wave case, we obtain
\begin{align}
	\braket{r_Q^2}^{(p)} &= \frac{3}{10} \frac{f^2}{m_R S_n^{(p)}}~,\\
	\braket{r_E^2}^{(s)} &= \frac{1}{4} \frac{f^2}{m_R S_n^{(s)}}~.
\end{align}
}

\begin{figure}[t]
\centering
\includegraphics[width=0.32\textwidth]{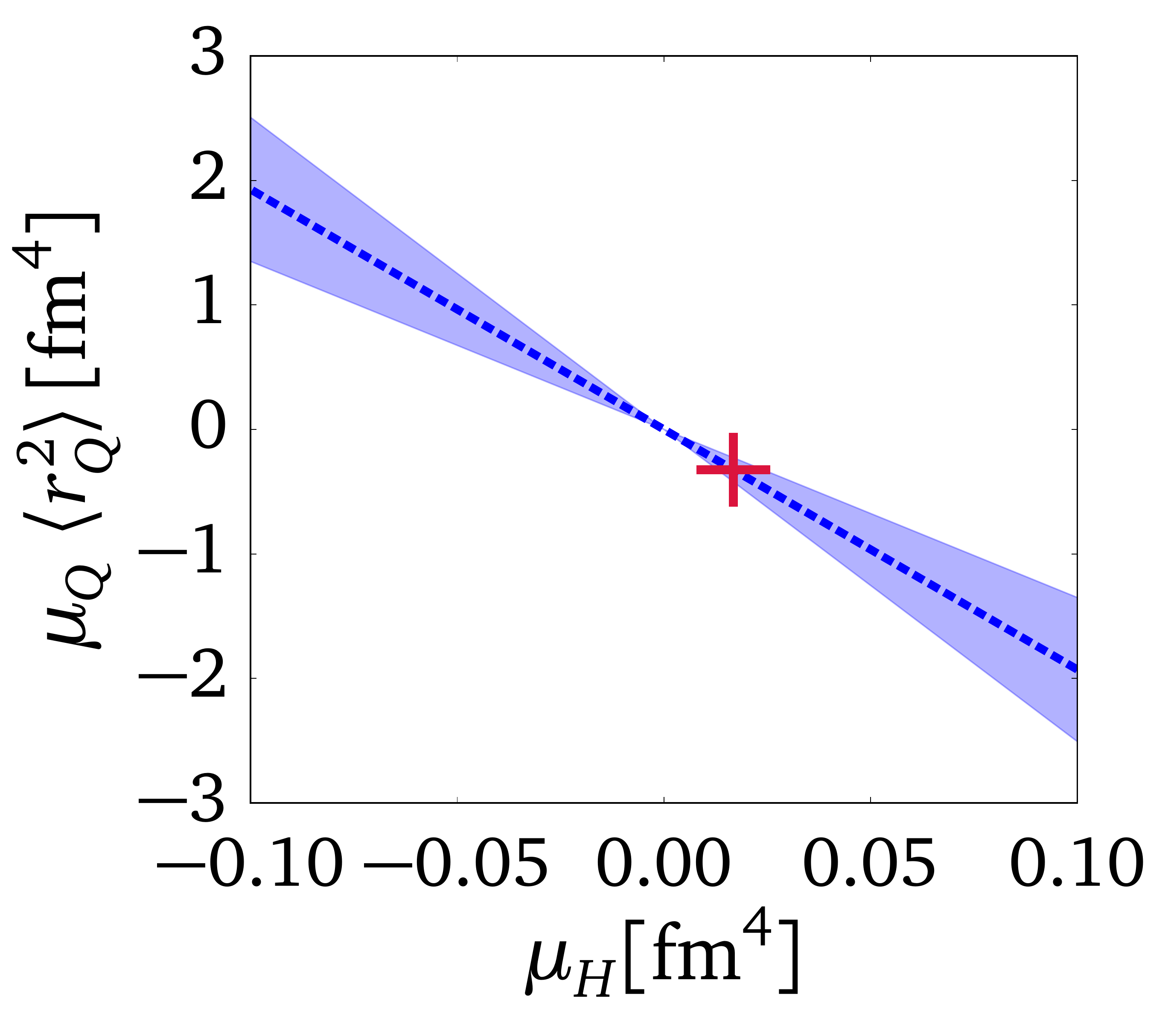}
\includegraphics[width=0.32\textwidth]{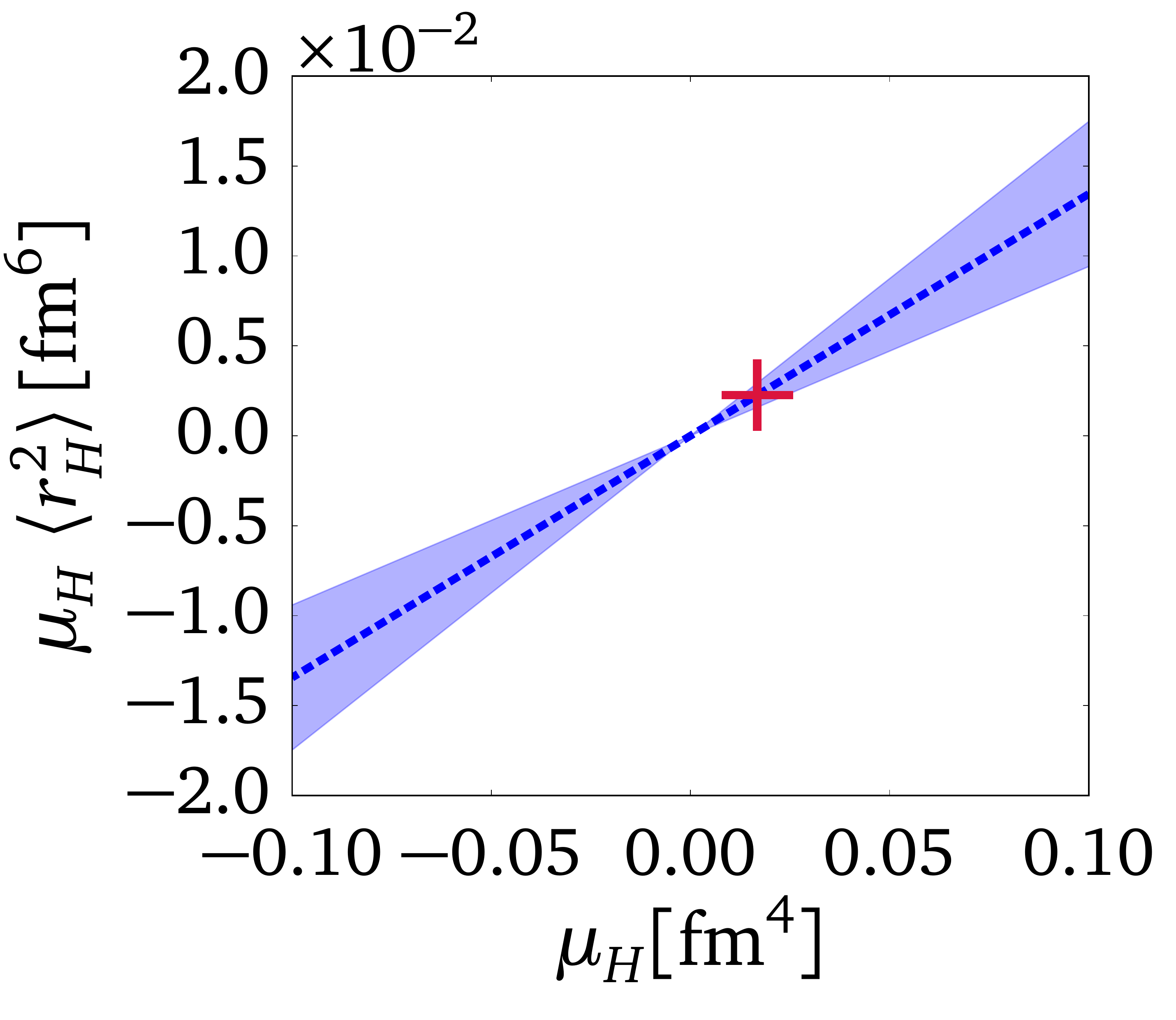} \\
\includegraphics[width=0.31\textwidth]{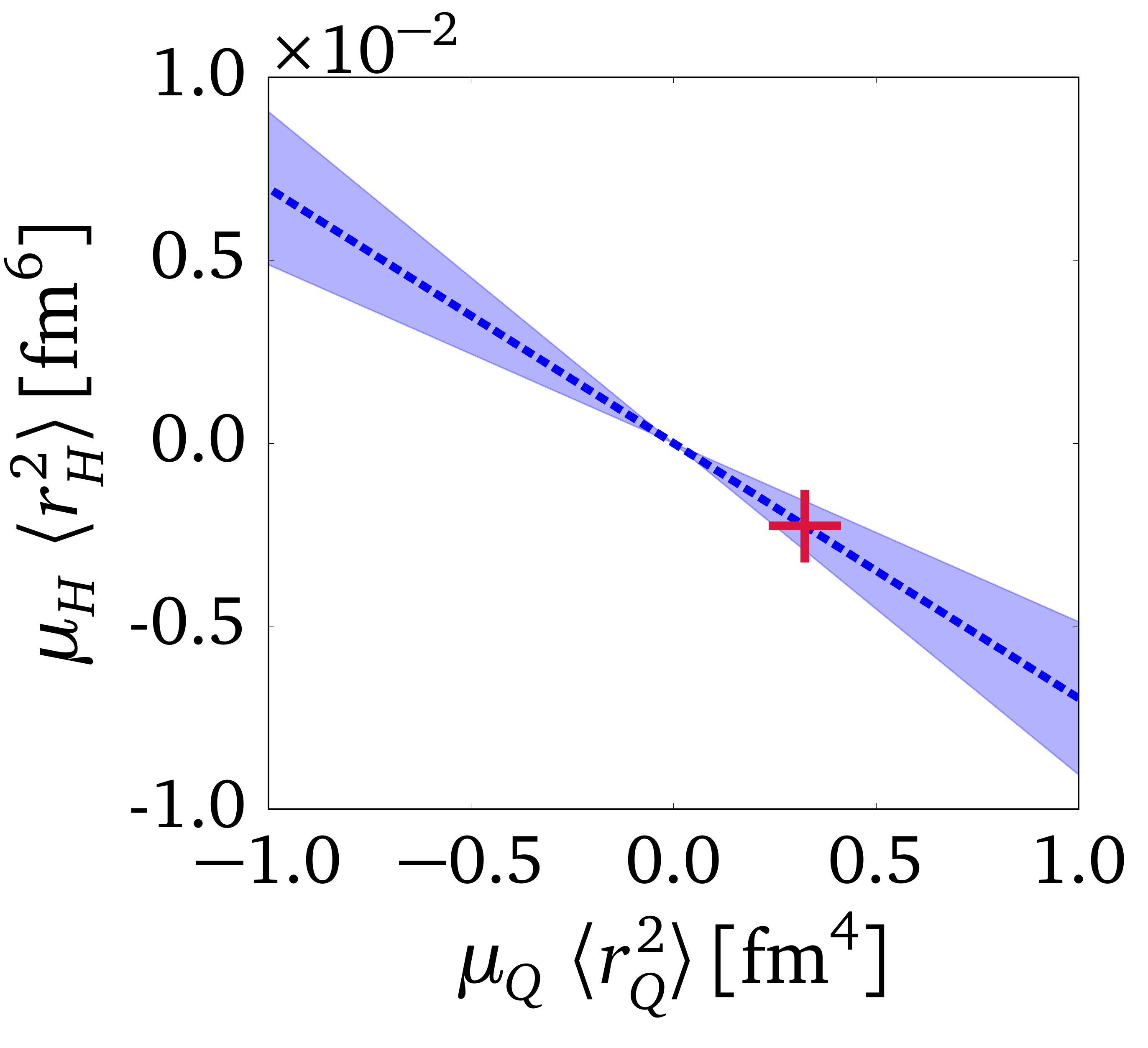}
\includegraphics[width=0.31\textwidth]{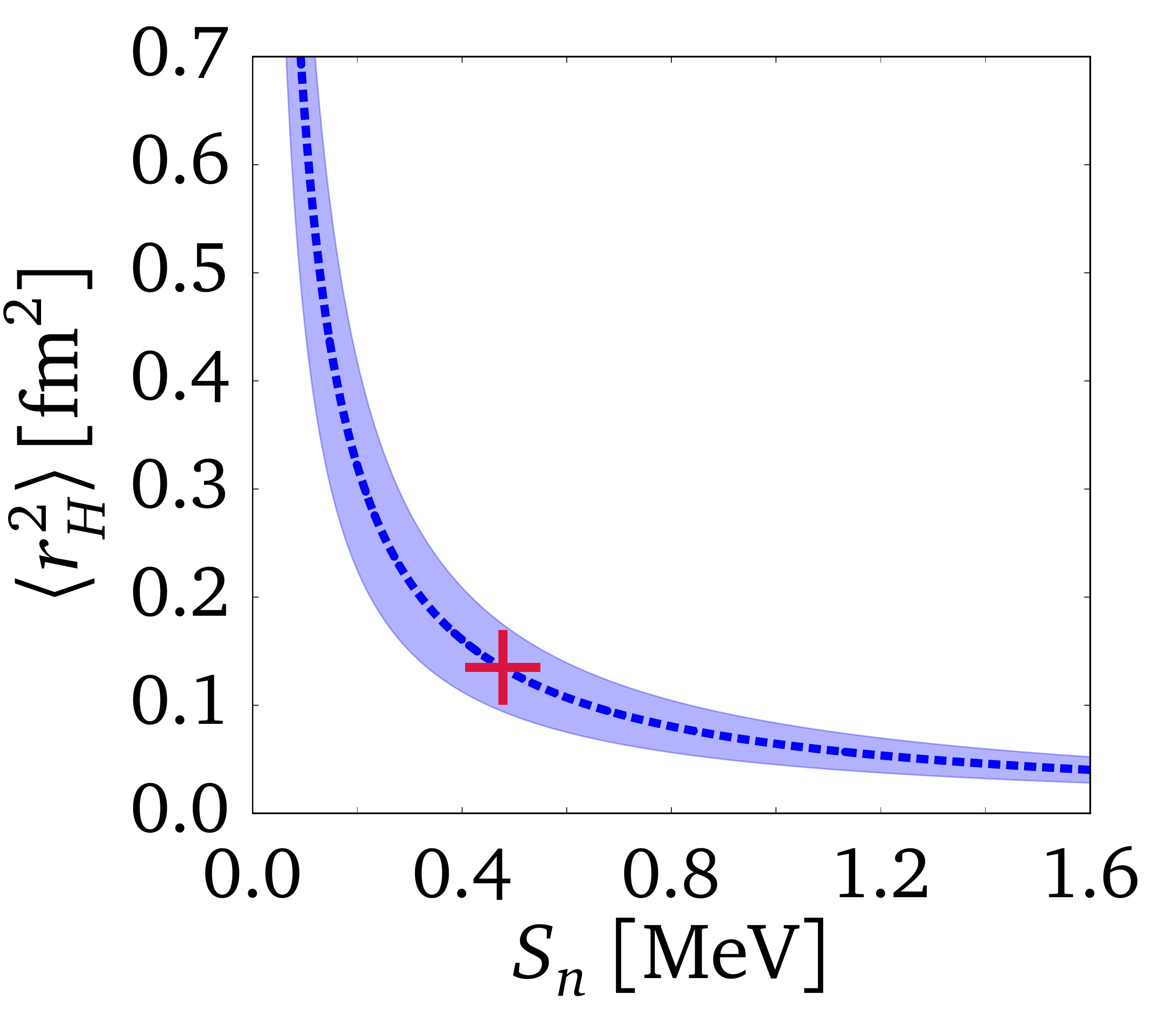}
\caption{Linear correlations between the hexadecapole moment and the quadrupole moment times quadrupole radius (top left), the hexadecapole moment and the hexadecapole moment times hexadecapole radius (top right) and between the quadrupole moment times quadrupole radius and hexadecapole moment times hexadecapole radius (bottom left). Bottom right: correlation between the neutron separation energy and the hexadecapole radius. The red cross denotes the numerical prediction for $^{15}$C. The EFT uncertainties are given by the shaded bands.}
\label{fig:correlations}
\end{figure}

For the $D$-wave, we can derive several linear correlations between different combinations of multipole 
moments and radii. This is illustrated in Fig.~\ref{fig:correlations}, where 
the red cross denotes the numerical prediction of the corresponding quantity 
for $^{15}$C. Therefore, by measuring one of these observables, we can 
immediately predict the correlated quantity. These correlations are universal 
and can be found in arbitrary one-neutron $D$-wave halo nuclei or similar 
weakly-bound systems.
%%%%%%%%%%%%%%%%%%%%%%%%%%%%%%%%%%%%%%%%%%%%%%%%%%%%%%%%%%%%%%%%%%%%%%%%%
%%%%%%%%%%%%%%%%%%%%%%%%%%%%%%%%%%%%%%%%%%%%%%%%%%%%%%%%%%%%%%%%%%%%%%%%%

With the numerical result for $Z_d m_R^2 g_2^2$, we can check if our 
power counting scenario, leading to the scaling $Z_d m_R^2 g_2^2 \sim 
R_{halo}^2 
R_{core}$, can be confirmed or if the scenario of~\cite{bedaque2003narrow} 
yields better agreement.
An approximation for the halo scale can be extracted from the neutron 
separation 
energy $S_n$, $R_{halo} \approx 1/\gamma_2 = 1/\sqrt{S_n 2 m_R} = 6.81 \text{ 
fm}$.
We can approximate the core scale by looking at the energy of the first 
excitation of the $^{14}$C nucleus $E_{ex} = 6.1$ MeV. Converting this energy 
into a length scale, we obtain $R_{core} \approx 1.91$ fm.
By employing the experimental values for $R_{halo}$ and $R_{core}$, we predict
$Z_d m_R^2 g_2^2 \sim R_{halo}^2 R_{core} \approx 90 \text{ fm}^3$. This value 
is only by a factor of $2$ smaller than the one extracted from B(E2) and considering
that this is an estimation grounded solely on the scaling within our power counting, our result is 
in reasonable agreement.
The power counting of~\cite{bedaque2003narrow} does lead to the scaling
$Z_d m_R^2 g_2^2 = 1/r_2 \sim R_{core}^3 \approx 7 \text{ fm}^3$ which is 
around 
$26$ times
smaller than the extracted result.
These numbers indicate that our power counting scenario is better suited
for $^{15}$C.

To obtain the correlation between the quadrupole transition
from the $\frac{5}{2}^+$ to the $\frac{1}{2}^+$ state and the
quadrupole moment of the $\frac{5}{2}^+$ state,
we combine Eqs.~\eqref{eq:NLO_muQ} and \eqref{eq:E2_LO} and apply a factor
2/6 to account for the different multiplicity of initial and final states.
We obtain a linear dependence between B(E2) for the transition
$\frac{5}{2}^+ \to\frac{1}{2}^+$ and the quadrupole moment
\begin{align}
\label{eq:haloCorr}
\text{B(E2)} = \frac{-1}{50 \pi} \ \frac{Z_\text{eff}^2 \, e^2 \gamma_0}
{(1-r_0\gamma_0)} \ \left[ \frac{3\gamma_0^2 + 9\gamma_0\gamma_2
+ 8\gamma_2^2}{(\gamma_0 + \gamma_2)^3} \right]^2
\frac{\mu_Q^{(d)}}{\tilde{L}_{C02}^{(d)} -\frac{5}{4} \gamma_2 f^2}~,
\end{align}
where $\tilde{L}_{C02}^{(d)}$ is treated as fit parameter and $\gamma_0$ and 
$\gamma_2$ are taken from experiment \cite{ajzenberg1991energy}.

%%%%%%%%%%%%%%%%%%%%%%%%%%%%%%%%%%%%%%%%%%%%%%%%%%%%%%%%%%%%%%%%%
% rigid rotor
%%%%%%%%%%%%%%%%%%%%%%%%%%%%%%%%%%%%%%%%%%%%%%%%%%%%%%%%%%%%%%%%%
A similar correlation between the quadrupole transition and the
quadrupole moment can be obtained from the rotational model by Bohr
and Mottelson~\cite{bohr1975nuclear}
\begin{align}
\label{eq:RigidRotor}
\text{B(E2}, J_i \rightarrow J_f) &= \frac{5}{16 \pi} \frac{\left( 
(J+1)(2J+3)\right)^2}{\left(3K^2-J(J+1)\right)^2} \\
&\quad\times \left(\left. J_i K \ 2 0 \right\vert J_f K \right)^2 
\left(\frac{Q^{0,t}}{Q^{0,s}}\right)^2 \mu_Q(J)^2 \ ,\notag
\end{align}
where $K=1/2$ denotes the projection of the total angular momentum on the 
symmetry axis of the intrinsically deformed nucleus and $Q^{0,t}/Q^{0,s}$ is 
the ratio between intrinsic static ($s$) and transition ($t$) quadrupole moment 
in the rigid rotor model.
The idea to employ this simple model is motivated by observations of Calci
and Roth~\cite{calci2016sensitivities}, who found a robust correlation between 
this pair of quadrupole observables in \textit{ab initio} calculations for 
light nuclei. In the simple rigid rotor model the ratio $Q^{0,t}/Q^{0,s}$ is 
expected to be one. The results of Ref.~\cite{calci2016sensitivities} indicate 
that the correlation is robust as long as the ratio $Q^{0,t}/Q^{0,s}$ is 
treated as a fit parameter.
%%%%%%%%%%%%%%%%%%%%%%%%%%%%%%%%%%%%%%%%%%%%%%%%%%%%%%%%%%%%%%%%%%%%%%%%%%%%%
\begin{figure*}[t]
\centering
\includegraphics[width=0.45\textwidth]{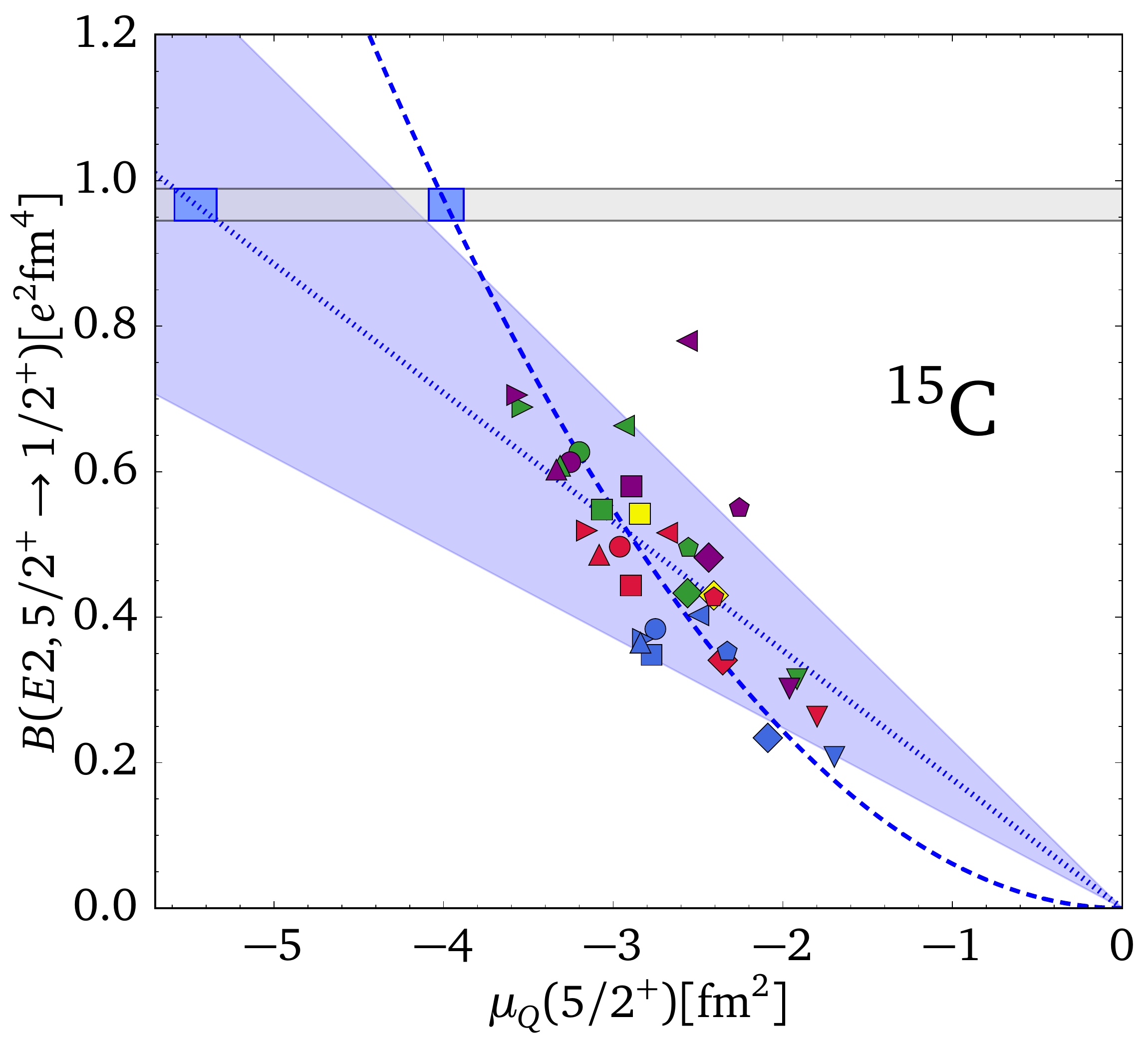}
\qquad
\includegraphics[width=0.45\textwidth]{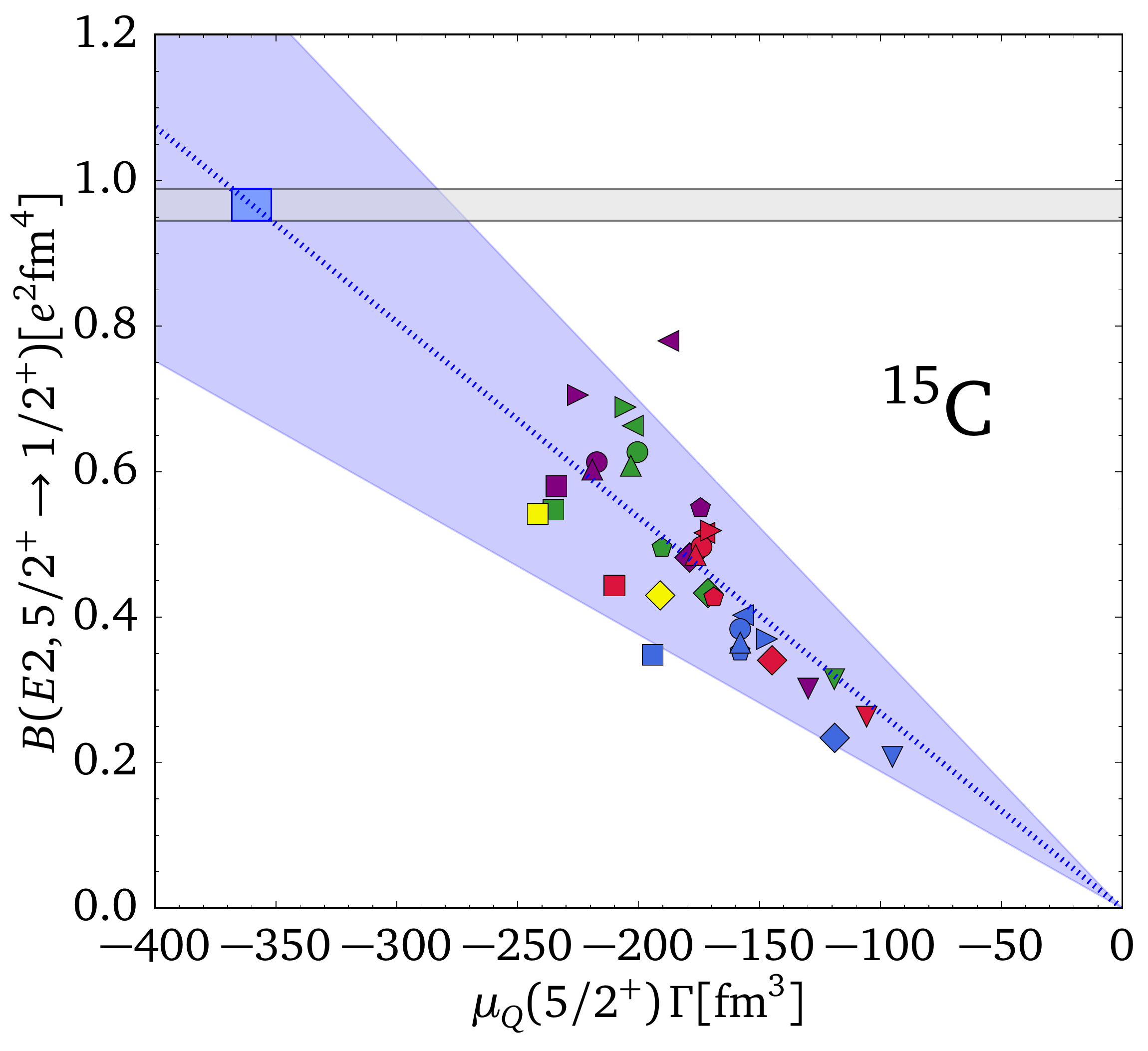}
\caption{Correlation between B(E2) and the quadrupole moment $\mu_Q$.
The IT-NCSM data is obtained with different NN+3N chiral EFT
interactions: EM with cutoffs $\{400,400,500\}$ MeV/c (square, diamond,
triangle down), and EGM with cutoffs $(\Lambda_\chi/\tilde{\Lambda}_\chi)
= \{(450/500),(600/500),(550/600),(450/700),(600/700)\}$ MeV/c (triangle
left, pentagon, circle, triangle right, and triangle up) with oscillator
frequency $\hbar \Omega = 16$ MeV for all IT-NCSM calculations except for
the diamond and triangle down data where $\hbar \Omega = 20$ MeV.
Different colors denote different $N_\text{max} = 2$ (blue), 4 (red), 6 (green),
8 (violet), and 10 (yellow) values.
Left panel: Rigid rotor model with quadratic fit of $Q^{0,t}/Q^{0,s}$
ratio (dashed
line, $\chi_{red}^2 = 110$) and linear Halo EFT fit of $\tilde{L}_{C02}^{(d)}$
with fixed $\gamma_2$ from experiment (dotted line, $\chi_{red}^2 = 123$).
Right panel: Linear Halo EFT fit with $\gamma_0^2-\gamma_2^2$ from IT-NCSM
calculation and rescaled $\mu_Q/\Gamma$ (dotted line, $\chi_{red}^2 = 80$),
where $\Gamma=\gamma_0 (3\gamma_0^2 + 9\gamma_0\gamma_2
+ 8\gamma_2^2)^2/(1-r_0\gamma_0)/(\gamma_0 + \gamma_2)^6$
divides out dependence on $\gamma_0$ and $\gamma_2$.
The gray shaded area indicates the uncertainty band of the experimental
B(E2)~\cite{ajzenberg1991energy}. The blue box within the gray shaded area corresponds to
the prediction for $\mu_Q$.
The EFT uncertainties are given by the blue shaded bands.}
\label{fig:C15dataAll}
\end{figure*}
%%%%%%%%%%%%%%%%%%%%%%%%%%%%%%%%%%%%%%%%%%%%%%%%%%%%%%%%%%%%%%%%%%%%%%%%%%%%%

We use IT-NCSM data of $^{15}$C, generated by different chiral EFT
interactions and different model spaces, to check the quadratic and
linear correlations and predict numerically the quadrupole moment of
$^{15}$C. This is demonstrated in Fig.~\ref{fig:C15dataAll}.
The varying symbols denote different \mbox{NN+3N} chiral EFT interactions which
are similar to the ones used in Ref.~\cite{calci2016sensitivities}.
We use the NN interaction developed by Entem and Machleidt 
(EM)~\cite{entem2003accurate} at N$^3$LO with a cutoff of 500 MeV/c for the 
nonlocal regulator function. This NN force is combined with 
the local 3N force 
at N$^2$LO using a cutoff of 400 or 500 MeV/c~\cite{navratil2007local}.
The second NN interaction by Epelbaum, Gl\"ockle, Mei{\ss}ner
(EGM)~\cite{epelbaum2004improving} at N$^2$LO uses a nonlocal regularization
with a cutoff $\Lambda_\chi$ and an additional spectral function regularization 
with cutoff $\tilde{\Lambda}_\chi$.
The EGM NN
forces are combined with a consistent nonlocal 3N force at N$^2$LO used in
several applications to neutron
matter~\cite{hebeler2013neutron,tews2013neutron,kruger2013neutron}.
For reasons of convergence, the NN+3N potentials are softened by a
similarity renormalization group evolution where all contributions up to the
three-body level are included.

We note that these interactions are based on Weinberg's power counting
\cite{Weinberg:1991um}
and their cutoff cannot be varied over a large range. 
However, chiral potentials based on Weinberg's power counting
have been very successful phenomenologically in nuclear structure and are
currently the only potentials available that are well tested for
p-shell nuclei. In particular, N$^2$LO EGM interactions are still
the only fully consistent set of two- and three-body interactions
for which significant experience with structure calculations in the p-shell
exists. Similarly, there is much experience with the EM interactions
supplemented with (inconsistent) 3N forces. For this reason, we use these
older interactions in our analysis. We believe that this is not a limiting
factor of our analysis. After all, we are interested in universal
properties which must emerge from any interaction that has the correct
low-energy physics. 

The different colors in Fig.~\ref{fig:C15dataAll} denote different 
$N_\text{max}$
values. The EFT uncertainties for the linear correlation
are given by the blue shaded bands.
Since the IT-NCSM results are not fully converged
and the results differ for different $N_\text{max}$ values,
the ordering of the ground and first excited state is exchanged for some
data points. Leaving out the data sets with exchanged ordering does not
significantly improve the fit.
The plot on the left side employs the experimental values for the neutron
separation energy as input for $\gamma_0$ and $\gamma_2$. For the plot on
the right side, we use the excitation energy of the first excited state
from the IT-NCSM to
determine $\gamma_0^2-\gamma_2^2$ and for $\gamma_0$ we use the experimental
value.

We emphasize that in the {\it ab initio} calculations, both, the interactions
(including short distance physics) and the model spaces are varied. 
If the {\it ab initio} calculations were (i) fully converged and
(ii) all interactions and electric operators were unitarily equivalent at the $A$-body level,
they would fall on a single point. However, neither (i) nor (ii) is
the case here. So, naively, one would
expect the calculations for B(E2) and $\mu_Q$ to fill the whole plane.
Halo EFT and the rotational model, however, predict a one parameter
correlation between B(E2) and $\mu_Q$ based on certain assumptions.
If these
assumptions, such as shallow binding and a corresponding separation
of scales in the case of Halo EFT,
are satisfied in the {\it ab initio} calculations, they should also
show the correlation even if they are not converged and/or have different
short distance physics. A similar behavior is observed in the case of the
Phillips and Tjon line correlations in light nuclei which are also
satisfied by "unphysical" calculations (See, e.g., Ref.~\cite{Nogga:2004ab}
for an explicit example).

An additional complication here is the appearance of the two-body
coupling $L_{C02}^{(d)}$ in
Eq.~(\ref{eq:localOperators}) which could vary for the different {\it ab
initio} data sets. In our analysis of the {\it ab initio} data, we explicitly
assume that $L_{C02}^{(d)}$ varies only slowly and can be approximated by a
constant for the {\it ab initio} data considered.\footnote{
A similar assumption
is made in the analysis of three-body recombination rates for ultracold
atoms near a Feshbach resonance to observe the Efimov effect. There the
scattering length varies strongly with the magnetic field B while the
three-body parameter is assumed to stay approximately
constant~\cite{Braaten:2004rn}.
Since the two parameters are independent it would be very unnatural if
both had a resonance at the same value of B.}
Under this assumption, it becomes possible to decide between the type of
correlation using the {\it ab initio} data for $^{15}$C.

From the left plot, we obtain $\mu_Q^{(d)} \approx -3.98 (5)$ fm$^{2}$
for the quadratic
fit and $\mu_Q^{(d)} \approx -5.46 (12)(1.64)$ fm$^{2}$
for the linear fit, where the uncertainties from B(E2)
are given in parenthesis. The second uncertainty for the linear fit
is from higher orders in the EFT.
From the fits, we cannot decide which scenario describes the IT-NCSM data
more appropriately since both lead to similar reduced $\chi^2$ values of $\chi_{red}^{2}=110$ for 
the quadratic fit and $\chi_{red}^{2}=123$ for the linear fit.
Please note that the absolute $\chi_{red}^2$ values have no significance
since the theoretical errors of the {\it ab initio} results were not included
in the fit, and only a relative comparison makes sense.
The ratio $Q^{0,t}/Q^{0,s}$ should be equal to $1$ for an ideal rigid rotor.
Since the quadratic
fit yields a ratio of $Q^{0,t}/Q^{0,s} \approx 0.5$, we assume that $^{15}$C is
not a good example of a rigid rotor.
Perhaps for larger $N_\text{max}$ values, and thus better converged results, the
matching between fit curves and data points would improve.

In the linear case, the slope of the fit depends also on the neutron
separation energies of both states, which differ for each data point
from the IT-NCSM. From the excitation energy obtained in the
IT-NCSM calculation, we only know
the difference between the neutron separation energies of the ground
and excited state. Thus, one experimental input is still required to fix
$\gamma_0$ and $\gamma_2$ from the IT-NCSM data, since we did not perform
explicit calculations for $^{14}$C.
In the right plot of Fig.~\ref{fig:C15dataAll}, we determine
$\gamma_0^2-\gamma_2^2$ from IT-NCSM data and take $\gamma_0$
from experiment. We deem this analysis to be more consistent
than the previous one.
The reduced $\chi^2$ value for
the linear fit then slightly improves to $\chi_{red}^{2}=80$ compared to the fit using
experimental values only.
This leads to
$\mu_Q^{(d)} \approx -4.21 (10)(1.26) \ \mathrm{fm}^2$, which is closer to the
value from the quadratic fit.
The deviations of the data points from the linear fit might decrease
further if consistent values for both neutron separation energies
were extracted from the IT-NCSM.
This NLO correlation is expected to hold up to
corrections of order $R_{core}/R_{halo} \approx 0.3$ given by the blue shaded band.
Taking this EFT uncertainty into account, the \textit{ab initio} data satisfy the 
correlation very well.

With the extracted results of $\mu_Q^{(d)}$, we can predict the quadrupole radius, $\braket{r_Q^2}^{(d)} = 5.93(13)(1.78)$ $\times$ $10^{-2}$ fm$^2$ from the left linear fit and $\braket{r_Q^2}^{(d)} = 7.70(17)(2.31) \times 10^{-2}$ fm$^2$ from the right linear fit in Fig.~\ref{fig:C15dataAll}, by Halo EFT.

Finally, we note that the NCSM calculations
for small $N_{max}$ are not converged in the IR. However, it can be clearly seen 
in Fig.~\ref{fig:C15dataAll}
that our conclusions are unchanged if the smallest
$N_{max}=2,4,6$ are omitted. In fact, Calci et al.~\cite{Calci:2016tvb}
showed explicitly
that the universal correlation between the B(E2: $0^+ \to 2^+$) and the
quadrupole moment of the $2^+$ state in $^{12}$C is extremely well
satisfied even for the smallest $N_{max}$.

\section{Conclusion}
\label{sec:conclusion}
We have extended the Halo EFT approach for 
electric observables
to shallow $D$-wave bound states. 
Additionally, a basic framework for the extension of our Halo EFT to higher 
partial waves has been outlined. We have developed
a power counting scheme for arbitrary $l$-th partial 
wave shallow bound states that differs from the scenario
of~\cite{bedaque2003narrow} for $l>1$. This power counting
was applied to $^{17}$C in Ref.~\cite{Braun:2018vez} where also some magnetic 
observables were considered. For higher partial waves the number of 
fine-tuned parameters increases.
Based on the assumption that a larger number of fine tunings is
less natural, this suggests that shallow bound states in higher partial waves 
are less likely than in lower ones, which is also
observed experimentally.

Using this scheme, we have computed the B(E2) strength at LO and found that no 
additional counterterm is required at this order.
We have also calculated the electric quadrupole as well as hexadecapole form 
factors at NLO and found a smooth, universal correlation between the quadrupole 
radius and the hexadecapole moment.
We find that for the $D$-wave, the local gauge invariant operators become
more important than in lower partial waves
and counterterms are required for the form factors already at NLO. This
continues the trend, observed in~\cite{hammer2011electric},
that the counterterms enter in lower orders at larger $l$.
The emergence of counterterms in low orders limits the predictive power of Halo 
EFT for $D$-waves. However, this limitation can be overcome by considering 
universal correlations between observables as discussed below.

We emphasize that, up to this point, all our results are universal and
not specific for $^{15}$C.
Considering now $^{15}$C as an example, the lack of data for
the first excited $\frac{5}{2}^+$ state makes numerical predictions
difficult.
Using our result for the B(E2) and by comparing it to the measured B(E2) data, 
we have been able to make predictions for the hexadecapole moment 
$\mu_H^{(d)} = 1.68(4)(50) 
\times 10^{-2}\ \text{fm}^{4}$ and radius $\braket{r_H^2}^{(d)} = 0.135(3)(40) \ \text{fm}^{2}$.
We cannot directly predict values for the charge radius and quadrupole
moment and radius at NLO since the expressions \eqref{eq:r_E_LO}, \eqref{eq:NLO_muQ} and \eqref{eq:rQ}
contain unknown counterterms.
Nevertheless, we have determined a value for the quadrupole moment,
$\mu_Q^{(d)} \approx -4.21 (10)(1.26) \ \mathrm{fm}^2$,
by exploiting the linear correlation between
the reduced E2 transition strength B(E2)
and the quadrupole moment in our Halo EFT and fitting the
unknown counterterm to \textit{ab initio} results from the IT-NCSM. For consistency reasons, 
we prefer the
result from the right plot of Fig.~\ref{fig:C15dataAll} using the excitation energy from IT-NCSM 
calculation.
With this result for the quadrupole moment, we have also predicted the quadrupole radius for 
$^{15}$C, $\braket{r_Q^2}^{(d)} \approx 7.70(17)(2.31)\times 10^{-2} \ \mathrm{fm}^2$, using universal correlations 
from Halo EFT.
These correlations are not
obvious in \textit{ab initio} approaches, since the separation of scales
is not explicit in the parameters of the theory. This demonstrates the
complementary character of Halo EFT towards \textit{ab initio} methods.
In principle, the universal correlations allow to extract information
even from unconverged \textit{ab initio} calculations since the
correlations are universal.
We have compared the linear Halo EFT correlation to the quadratic
correlation based on the simple rotational model by Bohr and Mottelson.
The value for the quadrupole moment,
$\mu_Q^{(d)} \approx -3.98 (5)\ \mathrm{fm}^2$,
obtained from the quadratic correlation deviates from the linear result
by 5\% -- 30\% depending on the input used for $\gamma_0^2-\gamma_2^2$.

While there is a clear correlation in the \textit{ab initio} data,
there are also some outliers.
In the case of the linear Halo EFT correlation, this could be
due to the use of the experimental value of the
ground state neutron separation energy $\gamma_0$, which is
presumably inconsistent with some of the \textit{ab initio} data sets.
Since the Halo EFT correlation depends on the exact neutron separation
energy of the two states, consistent values should be used.
However, within the EFT uncertainty the predicted correlation 
is well satisfied.
Better converged data sets and
the future determination of the neutron separation energy directly from the
IT-NCSM would help to clarify the situation.
This proves the usefulness of our Halo EFT approach even for $D$-wave bound 
states, but also demonstrates the limiting factors for the extension to higher 
partial waves.

\begin{acknowledgments}
We thank D.R. Phillips and L. Platter for discussions.
This work has been
funded by the Deutsche Forschungsgemeinschaft (DFG, German Research
Foundation) -- Projektnummer 279384907 -- SFB 1245
and by the Bundesministerium f\"ur Bildung und Forschung (BMBF)
through contract no. 05P18RDFN1.
\end{acknowledgments}

\appendix

\section{\textit{S}-wave propagator}
\label{app:Swave}
The dressed $\sigma$ propagator and the $S$-wave scattering amplitude are 
computed by summing the bubble diagrams analog to the $D$-wave case shown
in Fig.~\ref{fig:dimer-propagator}.
The result for the dressed $\sigma$ propagator is
\begin{align}
D_\sigma(p) &= \frac{1}{\Delta_0 + \eta_0 [p_0 - \mathbf{p}^2/(2 M_{nc}) 
+i\epsilon] -\Sigma_\sigma(p)}~,\\[6pt]
\Sigma_\sigma(p) &= - \frac{g^2_0 m_R}{2 \pi} \left[i \sqrt{2 m_R \left(p_0 - 
\frac{\mathbf{p}^2}{2 M_{nc}} \right)} + \mu\right]~,
\end{align}
where PDS is employed as regularization scheme with scale 
$\mu$~\cite{kaplan1998new, kaplan1998two}.
After matching to the effective range expansion, we obtain for the $\sigma$ 
propagator
\begin{align}
\notag
D_\sigma(p) &= Z_\sigma \frac{1}{p_0-\frac{\mathbf{p}^2}{2 M_{nc}}+B_0} + 
R_\sigma(p) \ ,
\end{align}
with
% \begin{align}
% & Z_\sigma = \frac{2\pi\gamma_0}{m_R^2 g_0^2}\quad \mbox{(LO)}\ , \quad
% Z_\sigma = \frac{2\pi }{m_R^2 g_0^2} \frac{\gamma_0}{1 -r_0 \gamma_0}\quad 
% \mbox{(NLO)}\ .
% \end{align}
\begin{align}
\label{eq:zs}
& Z_\sigma = \frac{2\pi\gamma_0}{m_R^2 g_0^2}\quad \mbox{(LO)}\ , \quad
Z_\sigma = \frac{2\pi }{m_R^2 g_0^2} \gamma_0 \left[1+\gamma_0 r_0\right] \quad 
\mbox{(NLO)}\ .
\end{align}
Here, $Z_\sigma$ denotes the wave-function renormalization, 
$B_0=\gamma_0^2/(2m_R)$ denotes the binding energy and the remainder 
$R_\sigma(p)$ is regular at the pole.

\bibliography{DWavePaper_V17.bbl}

\end{document}